\documentclass[aps,usenames,dvipsnames,
               prb,
               twocolumn,
               10pt,
               longbibliography,
               floatfix,
               ]{revtex4-2}

\usepackage{soul}
\usepackage{physics}
\usepackage{amssymb}
\usepackage{amsmath}
\usepackage[english]{babel}
\usepackage[utf8]{inputenc}
\usepackage[pdftex]{graphicx}
\usepackage{xspace}
\usepackage{dsfont}
\usepackage{bm}
\usepackage{comment}
\usepackage{float}

\usepackage[usenames, dvipsnames]{color}

\graphicspath{{.}{Figuras/}{Figures/}}

\usepackage{hyperref}
\hypersetup{
    colorlinks,%
    citecolor=blue,%
    filecolor=blue,%
    linkcolor=blue,%
    urlcolor=blue
}

\newcommand{\kp}{\ensuremath{\bm{k}\cdot\bm{p}}\xspace}

\begin{document}

\title{Physical Pictures for Quasisymmetry in Crystals}

\author{Bryan D. Assun\c{c}\~ao}
\author{Emmanuel V. C. Lopes}
\author{Tome M. Schmidt}
\author{Gerson J. Ferreira}
\affiliation{Instituto de F\'{i}sica, Universidade Federal de Uberl\^{a}ndia, Uberl\^{a}ndia, Minas Gerais 38400-902, Brazil}

\date{\today}

\begin{abstract}
Quasisymmetry (QS) provides a novel route to understand and control near-degeneracies, Berry curvature, optical selection rules, and symmetry-protected phenomena in quantum materials. Here we give physical interpretations of the emergence of QS operators across multiple material families. Using density functional theory and the \kp formalism, we identify QS subspaces and calculate their representation matrices, quantifying the quasisymmetry via a metric $\epsilon$ that measures subspace invariance. For Sn/SiC and transition-metal dichalcogenide monolayers, QS corresponds to an emergent mirror symmetry, whereas in wurtzite crystals it manifests as an emergent spatial inversion. By contrast, for AgLa the QS appearing in avoided crossings is inherited from a nearby high-symmetry point rather than being an emergent lattice symmetry. Combining group-theoretical analysis and \kp modeling, our results establish concrete physical pictures for QS and provide practical criteria to diagnose it in first-principles calculations.
\end{abstract}
\maketitle

\section{Introduction}

Symmetry is fundamental to condensed matter physics, as it governs key properties such as electronic band structure, optical selection rules \cite{koster1963properties, bir1974symmetry, winkler2003spin, Dresselhaus2007, willatzen2009kp, cardona2010fundamentals}, and topological classification \cite{AltlandZirnbauer1997TenFoldWay, Slager2012, Chiu2016Classification, Kruthoff2017, bradlyn2017topological, cano2018building, Cano2021}. These principles have guided the discovery and engineering of advanced quantum materials, including topological (crystalline) insulators \cite{KaneMele2005SHEGraphene, KaneMele2005Z2QSHE, Bernevig2006, Konig2007, Hsieh2008, Hsieh2009, Hasan2010, Fu2011TCI, Tanaka2012, Dziawa2012}, Weyl and Dirac semimetals \cite{Wan2011WeylSemimetal, Burkov2011WeylSemimetal,Singh2012WeylSemimetal, Young2012DiracSemimetal, Armitage2018WeylDirac}, topological superconductors \cite{Hasan2017Review, Schnyder2008TopoSupercond, Kitaev2009, Sato2017, Neupert2021, Zhang2024}, and altermagnets \cite{Smejkal2022Altermagnetism, Smejkal2022EmergingAltermagnetism, Krempask2024Altermagnetic, Osumi2024Altermagnetic}.
Since the early days, effective models are derived from symmetry principles via the theory of invariants \cite{bir1974symmetry}, \kp \cite{willatzen2009kp} and tight-binding \cite{Slater1954, Goringe1997} methods. Seminal works have used these approaches to describe essential materials, including Si and Ge \cite{LuttingerKohn1955}, zincblende \cite{Kane1956, Kane1957, LuttingerKohn1955, Cardona1966}, and wurtzite 
\cite{Hopfdeld1960, Cardona1966, Voon1996, ChuangChang1996, Gutsche1967, pikus1962new} crystals. In the absence of inversion symmetry, symmetry constraints allow for Rashba and Dresselhaus spin-orbit couplings \cite{Rashba1984, Dresselhaus1955SOC, Acosta2021SOC}, which are crucial for spintronics~\cite{liu2019spintronic}. 
Symmetry constraints are also essential to the development of novel materials for next-generation electronics, optoelectronics and energy storage, such as transition-metal dichalcogenides (TMDs) \cite{Splendiani2010, Xiao2012, Geim2013, mak2010atomically, mccreary2016synthesis, Woniak2020, singh2020tunable, jones2013optical, FariaJr2025InterGFactor, Wang2012TMDReview, Chhowalla2013TMDReview, MakShan2016TMDReview, Manzeli2017TMDReview, Glazov2024, Deb2024, Han2018TMDReview}, Janus \cite{Yagmurcukardes2020janus, Zheng2024janus, Zhang2020JanusReview, Yin2021Janusreview}, perovskites \cite{Testelin2022gfactor, StranksSnaith2015PerovskiteReview, Jena2019PerovskiteReview}, and MXenes~\cite{Naguib2013MXenesreview, Khazaei2017MXenesreview, Guha2022MXenes} crystals.

Yet, many materials show near-degeneracies and selection rules beyond crystallographic symmetry constraints, pointing to approximate symmetries. In disordered topological materials, \textit{ensemble averages} restore broken symmetries and produce topological Anderson insulators \cite{Li2009, Groth2009, Sttzer2018, Zhang2022}, for which the ensemble average of local invariants characterizes the topological phase \cite{Caio2019, Varjas2020, Bryan2024TAI}. Reference \cite{kitamura2018spinhalleffect2d} introduces the concept of vicinity topology in $\text{PtCoO}_2$, where an approximate mirror symmetry yields an emergent mirror Chern metal phase.
In perovskites, ferroelectrics and multiferroics, \textit{pseudosymmetries} occur when the crystal lattice is close to a higher-symmetry structure \cite{Kroumova2001, Zwart2007, capillas2011new, Nolze2023pseudo}. Particularly, in the \textit{quasicubic} approximation \cite{Hopfdeld1960, Cardona1966, Voon1996, ChuangChang1996, Gutsche1967}, wurtzite crystals are approximated by a cubic zincblende structure strained along the (111) direction. 
Orbital-dependent site-permutation symmetries \cite{Felipe2020SitePermutation} can also extend beyond the crystal group and protect higher band degeneracies.

The recent proposal of \textit{quasisymmetries} (QS) \cite{Guo2022, Hu2023Hierarchy, PRL2024} provides a framework to understand approximate symmetries in quantum materials. In CoSi \cite{Guo2022, Hu2023Hierarchy}, QS emerges as an angular momentum algebra protecting near-degeneracies and enhancing Berry curvature, which is detectable via quantum oscillations. Later, a classification of emergent QS groups, within the 32 point groups, has been developed in Ref.~\cite{PRL2024}. Quasisymmetry has since been identified in near-degenerate SOC gaps \cite{Yaji2019, Tao2023}, unconventional spin-polarizations \cite{Tao2024NonSymmorphic}, quantum spin Hall physics and topological quadrupoles \cite{Liu2024, Liu2024_orbital, li2026quasisymmetry, Zhang2025}, altermagnetism \cite{Roig2025Altermagnet}, asymmetrical Majorana bound states \cite{Kang2024Majorna}, Ising-type SOC in superconductors without $M_z$ symmetry \cite{Tao2023_ising}, and periodic dynamics in nonintegrable Hamiltonians \cite{Ren2021, Ren2024}. This variety of topics highlights the relevance of QS mechanisms to the design of advanced quantum materials, which motivates the question: What physical mechanisms drive the emergence of quasisymmetry in real materials?

In this paper, we identify two distinct and nonequivalent physical pictures for quasisymmetries and verify them through first-principles calculations. In the first picture, the crystal group is augmented by an emergent symmetry arising from significant wavefunction localization on a sublattice. It occurs in Sn/SiC \cite{Yaji2019, Tao2023} and TMD monolayers as mirror symmetries, and in wurtzite crystals as spatial inversion. Additionally, we argue that the quasicubic approximation for wurtzites is a special case of this first picture, where the approximation relies on compatibility relations rather than an emergent symmetry. 
The second picture appears in AgLa \cite{PRL2024}. There, quasisymmetry at generic k-points arises from symmetry selection rules inherited from nearby high-symmetry points.
We calculate the band structures and representation matrices for the QS operators, identifying quasi-irreps and quantifying quasisymmetry via a metric $\epsilon$ that measures subspace invariance. We focus on spin-orbit coupling (SOC) as the perturbation, though results can be extended to others, such as strain \cite{PRL2024}, electric fields and optical transitions. Using Löwdin's perturbation theory \cite{Lwdin1951}, we compare the first and higher-order SOC gaps, verifying the QS suppression of the first-order term.

This paper is organized as follows. Section \ref{sec:Methods} presents the theoretical framework including quasisymmetry formalism, the \kp picture, the metric $\epsilon$, and the methodology to calculate the first-order SOC gap. Section \ref{sec:results} presents our results for various materials. Section \ref{sec:conclusions} summarizes conclusions. Throughout the paper, the DFT calculations were done with both \texttt{Quantum Espresso} \cite{Giannozzi2009, Giannozzi2017} and \texttt{VASP} \cite{Kresse1993, Kresse1996}, Wannier models were obtained with \texttt{wannier90} \cite{mostofi2014updated, pizzi2020wannier90, mostofi2008wannier90} and \texttt{PythTB} \cite{PythTB}, representation matrices, symmetry analysis and effective models were developed via \texttt{DFT2kp} \cite{Cassiano2024}, \texttt{IrRep} \cite{IrRep2022}, and \texttt{Qsymm} \cite{varjas2018qsymm}.
The computational details are presented in Appendixes \ref{sec:computational} and \ref{sec:SOC_kp}. 

\section{Quasisymmetries}
\label{sec:Methods}

Here we identify two distinct and nonequivalent physical pictures for quasisymmetry, corresponding to different underlying mechanisms. The emergent-symmetry picture, discussed in Sec.~\ref{sec:emergent}, arises from real-space localization of the wave functions, which enhances the symmetry within the projected subspace and can augment the crystal group following the protocol presented in Ref.~\cite{PRL2024}. In contrast, the inheritance picture, presented in Sec.~\ref{sec:kppicture}, originates from the proximity to high-symmetry points in reciprocal space and does not rely on orbital or spatial localization. Although both mechanisms suppress first-order matrix elements within a subspace, their physical origins are fundamentally different. Additionally, Sec.~\ref{sec:metric} introduces a metric $\epsilon$ to quantify quasisymmetry in the emergent symmetry picture, and Sec.~\ref{sec:SOCpert} presents a method to extract the SOC intensity from DFT data.

\subsection{Quasisymmetries I: Emergent symmetry picture}
\label{sec:emergent}

We briefly review the framework introduced in Ref.~\cite{PRL2024}, where quasisymmetry is described as additional symmetries acting within the first-order eigensubspace $A$ of Löwdin's partitioning \cite{Lwdin1951}, leading to an augmented quasisymmetry group.

Generically, the Löwdin partitioning expresses the effective Hamiltonian $H$ in terms of matrix elements of the unperturbed Hamiltonian $H^0$ and a perturbation $H'$ as 
\begin{align}
    \label{eq:lowdin}
    H_{m,n} &=
        E^0_n \delta_{m,n}
        + H^{(1)}_{m,n} + H^{(2)}_{m,n} + \cdots,
    \\
    H^{(1)}_{m,n} &= H'_{m,n},
    \\
    H^{(2)}_{m,n} &= \dfrac{1}{2} 
    \sum_{r \in B}
    H'_{m,r}H'_{r,n}
    \left(
        \dfrac{1}{E^0_m - E^0_r}
        +
        \dfrac{1}{E^0_n - E^0_r}
    \right),
\end{align}
where $E^0_n$ are the eigenenergies for each eigenstate $\ket{n}$ of $H^0$, and $H'_{m,n} = \bra{m}H'\ket{n}$ are the matrix elements of the perturbation. Here, $(m,n) \in A$ refer to the set $A$ of eigenstates that constitute the subspace we want to model, while the summation in the second-order terms runs over the complementary set $B$ of remote bands. The first-order term $H^{(1)}_{m,n}$ is entirely defined within $A$, while higher-order terms $H^{(n\geq 2)}_{m,n}$ require both sets $A\oplus B$.

\begin{figure}[t]
    \centering
    \includegraphics[width=\columnwidth]{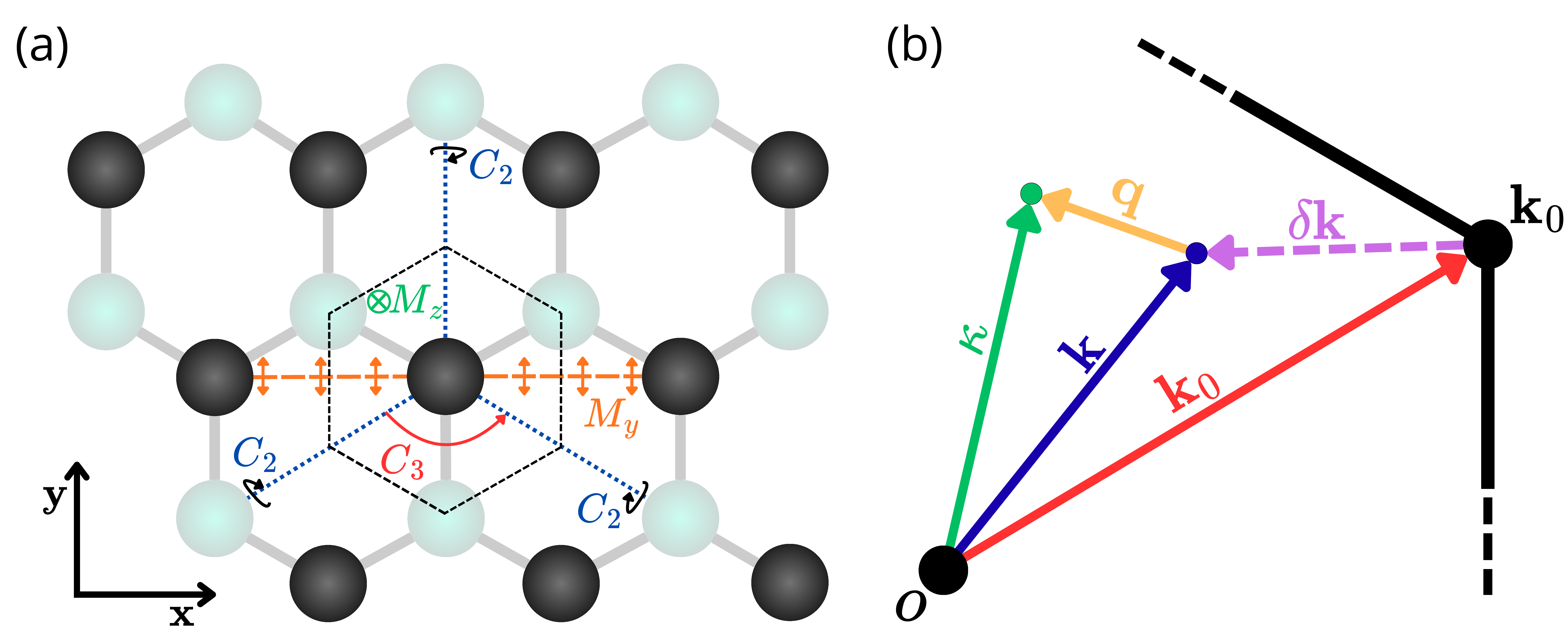}
    \caption{(a) Top view of a 2H transition metal dichalcogenide crystal lattice composed of two inequivalent sublattices: transition-metal sites (black), and chalcogen atoms (cyan). The full crystal group is $D_{3h}^1$, which is generated by $C_3(z)$, $C_2(y)$ and $M_z$, while the isolated transition-metal sublattice is invariant under an additional symmetry $M_y$ and transforms as the larger group $D_{6h}^1 = D_{3h}^1 \rtimes \{E, M_y\}$. (b) Schematic representation of the k-points used in the \kp expansion of the inheritance picture. The central point for the expansion is $\bm{k} = \bm{k}_0 + \delta\bm{k}$, which is near the high-symmetry point $\bm{k}_0$. The Bloch theorem is initially written for a generic point $\bm{\kappa} = \bm{k} + \bm{q}$, and later we consider $\bm{q}=0$.}
    \label{fig: kpoints}
\end{figure}

Within the group-theoretical framework, the presence or absence of first-order couplings $H^{(1)}_{m,n}$ is dictated by the symmetry constraints. At a generic point $\bm{k}$, its little group $\mathcal{G}_{\bm{k}}$ is a subgroup of the crystal group $\mathcal{G}_\Gamma$. Labeling the irreducible representations (irreps) of $\mathcal{G}_{\bm{k}}$ as $\Gamma^{(\bm{k})}_m$, the matrix element $H'_{m,n}$ can be finite if and only if $\Gamma^{(\bm{k})}_0 \subseteq \Gamma^{(\bm{k},*)}_m \otimes \Gamma^{(\bm{k})}_{H'} \otimes \Gamma^{(\bm{k})}_n$, where $\Gamma^{(\bm{k})}_0$ is the trivial irrep.

As shown in Ref.~\cite{PRL2024}, the basis functions for certain subspaces $A$ of $\mathcal{G}_{\bm{k}}$ are invariant under emergent quasisymmetries $Q = \{Q_1, Q_2, \dots\}$. They obey selection rules for an augmented quasisymmetry group $\mathcal{G}_Q = \mathcal{G}_{\bm{k}} \rtimes \{E, Q\}$ for which $\Gamma^{(Q)}_0 \not\subseteq \Gamma^{(Q,*)}_m \otimes \Gamma^{(Q)}_{H'} \otimes \Gamma^{(Q)}_n$. The quasisymmetry then imposes vanishing first-order terms $H^{(1)}_{m,n} = H'_{m,n} \equiv 0$, making the second-order term $H^{(2)}_{m,n}$ in Eq.~\eqref{eq:lowdin} the leading-order perturbation for $H_{m,n}$, where $H'_{m,r}$ obeys selection rules strictly from $\mathcal{G}_{\bm{k}}$ since $r \in B$.

This framework is insightful, but it does not provide an interpretation of why and when such quasisymmetries emerge. Here, we identify that the extra symmetries arise due to significant wave function localization on a single sublattice, which is invariant under an additional symmetry $Q$ that does not belong to the crystal group, as shown in Fig.~\ref{fig: kpoints}(a). The quasisymmetry $Q$ must be consistent with the little group $\mathcal{G}_{\bm{k}}$ of the k-point of interest to consistently generate the augmented group $\mathcal{G}_Q = \mathcal{G}_{\bm{k}} \rtimes Q$.

In Sec.~\ref{sec:results}, we show that for Sn/SiC \cite{Yaji2019, Tao2023} and monolayer 2H TMDs, the quasisymmetry corresponds to an in-plane mirror ($M_y$), while for wurtzites it is spatial inversion. Additionally, in Sec.~\ref{sec:Wurtzite} we show that the quasicubic approximation for wurtzites is a special case of this first picture, where the augmented group $\mathcal{G}_Q$ is built from compatibility relations rather than a direct product with an emergent symmetry $Q$.

\subsection{Quasisymmetries II: Inheritance picture}
\label{sec:kppicture}

The second quasisymmetry picture is due to the residual influence of a nearby high-symmetry point $\bm{k}_0$. Eigenstates at $\bm{k} = \bm{k}_0 + \delta \bm{k}$ approximately obey selection rules inherited from the nearby $\bm{k}_0$ point, little group $\mathcal{G}_{\bm{k}_0} > \mathcal{G}_{\bm{k}}$, when sufficiently isolated from remote bands.

To formulate this picture, we express the \kp framework with the Bloch theorem written as
\begin{align}
    \psi_{n,\bm{\kappa}}(\bm{r}) &= e^{i\bm{\kappa}\cdot\bm{r}} u_{n,\bm{\kappa}}(\bm{r}) = e^{i\bm{q}\cdot\bm{r}} \phi_{n,\bm{\kappa},\bm{q}}(\bm{r}),
    \\
    \phi_{n,\bm{\kappa},\bm{q}}(\bm{r}) &= e^{i(\bm{\kappa}-\bm{q})\cdot\bm{r}} u_{n,\bm{\kappa}}(\bm{r}),
    \label{eq:phi}
\end{align}
where $u_{n,\bm{\kappa}}(\bm{r})$ is the periodic part of the Bloch function $\psi_{n,\bm{\kappa}}(\bm{r})$.
Figure~\ref{fig: kpoints}(b) illustrates the k-points: $\bm{\kappa} = \bm{k} + \bm{q}$ is generic, $\bm{k}_0$ is a high-symmetry point, $\bm{k} = \bm{k}_0 + \delta\bm{k}$ is the central point of interest, and $\bm{q}$ is the deviation from $\bm{k}$. The \kp Hamiltonian acting on $\phi_{n,\bm{\kappa},\bm{q}}(\bm{r})$ is $H = H^0 + H^{\rm SOC} + 2 \bm{q}\cdot\bm{\pi}$, with
\begin{align}
    H^0 &= p^2 + V(\bm{r}) + q^2,
    \label{eq:H0}
    \\
    H^{\rm SOC} &= \bm{H}_\sigma \cdot\bm{\sigma},
    \\
    \bm{H}_\sigma &= \dfrac{\alpha^2}{4}(\nabla V\times\bm{p}).
    \label{eq:HsocSigma}
\end{align}
The orbital part $\bm{H}_\sigma = (H_x, H_y, H_z)$ of the spin-orbit coupling is defined to be used later, $V(\bm{r})$ is the crystalline potential, $\bm{\pi} = \bm{p} + \alpha^2 \bm{\sigma}\times\nabla V/8$ is the generalized momentum with SOC contributions, and $\alpha \approx 1/137$ is the fine structure constant. We use atomic Rydberg units, with $\hbar = 2m_0 = 1$. Hereafter, we focus on the central point of the \kp expansion $\bm{\kappa} \equiv \bm{k}$, with $\bm{q} = 0$.

Consider that two eigenstates $\ket{m, \bm{k}}$ and $\ket{n, \bm{k}}$ of $H^0$ cross at $\bm{k}$ with energies $E_m(\bm{k}) = E_n(\bm{k})$. Including SOC perturbatively, i.e., $H' = H^{\rm SOC}$, the first-order coupling to induce an anticrossing would be $\bra{m, \bm{k}}H'\ket{n, \bm{k}}$. To apply selection rules to this matrix element, we must consider the irreps of its constituents on the little group $\mathcal{G}_{\bm{k}}$, such that the matrix element is finite if and only if $\Gamma_m^{(\bm{k},*)} \otimes \Gamma_{H'}^{(\bm{k})} \otimes \Gamma_n^{(\bm{k})} \supseteq \Gamma_0^{(\bm{k})}$.

If $\bm{k} \approx \bm{k}_0$, we can approximate
\begin{equation}
    u_{n,\bm{k}}(\bm{r})
    \simeq
    u_{n,\bm{k}_0}(\bm{r})
    +
    \sum_{l \neq n}
    \frac{\delta\bm{k}\cdot \bm{p}_{l, n}}
    {\Delta_{n,l}}
    u_{l,\bm{k}_{0}}(\bm{r}),
\end{equation}
where both the energy difference $\Delta_{nl} = E_{n}(\bm{k}_{0})-E_{l}(\bm{k}_{0})$ and $\bm{p}_{l,n}= \bra{l,\bm{k}_0}\bm{p}\ket{n,\bm{k}_0}$ are evaluated at $\bm{k}_0$. 
Here, we assume that symmetry constraints $p_{m,n} = 0$ for the crossing states $\ket{m, \bm{k}}$ and $\ket{n, \bm{k}}$, such that only remote bands contribute to the second term above. In this case, the second term can be fully neglected if the crossing is sufficiently isolated from remote bands. Explicitly, the validity of the approximation is controlled by the condition $|\delta \bm{k} \cdot \bm{p}_{l,n}/\Delta_{n,l}| \ll 1$, which defines the regime where the inherited quasisymmetry holds.
Equation~\eqref{eq:phi} then becomes
\begin{equation}
    \phi_{n,\bm{k}}(\bm{r}) \equiv e^{i\bm{k}\cdot\bm{r}} u_{n,\bm{k}}(\bm{r}) \approx e^{i\delta\bm{k}\cdot\bm{r}}\phi_{n,\bm{k}_0}(\bm{r}),
    \label{eq:phase}
\end{equation}
where we now omit the $\bm{q}=0$ index for simplicity. Notice that $\phi_{n,\bm{k}_0}(\bm{r}) = e^{i\bm{k}_0\cdot\bm{r}} u_{n,\bm{k}_0}(\bm{r})$ is an eigenstate of $H^0$ at the high-symmetry point $\bm{k}_0$.

\begin{widetext}
    The approximation above allows us to write the matrix element of $H^{\rm SOC}$ as
    \begin{equation}
        \bra{m, \bm{k}}H'\ket{n,\bm{k}} 
            = \bra{m,\bm{k}_0}
                e^{-i\delta\bm{k}\cdot\bm{r}} H' e^{i\delta\bm{k}\cdot\bm{r}}
            \ket{n,\bm{k}_0}
            = \dfrac{\alpha^2}{4} \bra{m,\bm{k}_0} \Big[\nabla V \times (\bm{p}+\delta\bm{k}) \Big]\cdot \bm{\sigma} \ket{n,\bm{k}_0},
        \label{eq:SecondPicture}
    \end{equation}
    which can now be evaluated under the selection rules of $\mathcal{G}_{\bm{k}_0}$ for the high-symmetry point $\bm{k}_0$ instead of $\mathcal{G}_{\bm{k}} < \mathcal{G}_{\bm{k}_0}$. 
\end{widetext}

In AgLa \cite{PRL2024} (Sec.~\ref{sec:AgLa}), selection rules from $\mathcal{G}_{\bm{k}}$ allow for a finite $\bra{m, \bm{k}}H'\ket{n,\bm{k}}$, but stronger constraints from $\mathcal{G}_{\bm{k}_0}$ show that only the $\delta\bm{k}$-term in Eq.~\eqref{eq:SecondPicture} contributes, while the $\bm{p}$-term vanishes by symmetry. However, the $\delta\bm{k}$-term contribution to SOC is typically small and can be neglected. Thus, Eq.~\eqref{eq:SecondPicture} reduces to $\bra{m,\bm{k}}H'\ket{n,\bm{k}} = \bra{m,\bm{k}_0}H'\ket{n,\bm{k}_0}$. The left-hand side obeys selection rules from $\mathcal{G}_{\bm{k}}$, while the right-hand side obeys constraints from $\mathcal{G}_{\bm{k}_0}$. Therefore, all symmetries in $\mathcal{G}_{\bm{k}_0}$ that are not in $\mathcal{G}_{\bm{k}}$ act as inherited quasisymmetries at $\bm{k}$.

\begin{figure*}[t]
    \centering
    \includegraphics[width=\textwidth]{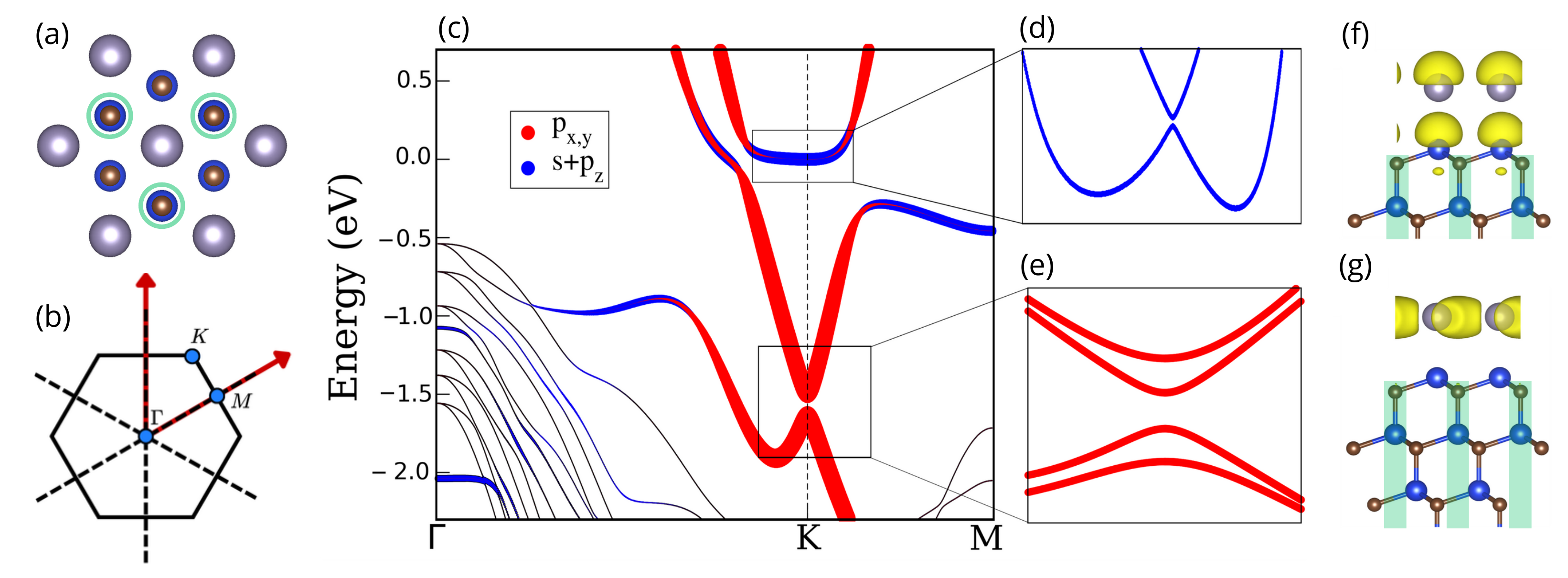}
    \caption{(a) Crystal structure of the Sn/SiC(0001)-$1\times$1 surface, where Sn atoms are adsorbed on top of surface Si atoms (on-top sites). Gray, blue, and brown spheres denote Sn, Si, and C atoms, respectively. The green contours highlight one of the two distinct Si-C stacking sequences not aligned with Sn atoms, as also illustrated in (f) and (g). (b) Corresponding Brillouin zone with high-symmetry points indicated. (c) Spin-orbit-free band structure projected onto Sn $s$ + $p_z$ (blue) and $p_x$ + $p_y$ (red) orbitals. Contributions from Si and C atoms are negligible in the energy window of interest and are therefore not shown. (d) Rashba-like (RL) spin splitting near the K point, with a SOC-induced gap of approximately 2 meV. (e) Zeeman-like (ZL) spin splitting near K, with a SOC-induced gap of approximately 157 meV. (f) and (g) Charge density distribution associated with the RL and ZL bands, respectively.}
    \label{fig:SnSiC}
\end{figure*}

\subsection{Quasisymmetry metric \texorpdfstring{$\epsilon$}{ϵ} and quasi-irreps}
\label{sec:metric}

We quantify quasisymmetry by measuring how closely a subset $A$ is invariant under a proposed QS operator $Q$. If $A$ is invariant under $Q$, the action of $Q$ is closed within $A$. In a block representation of $Q$ in the basis $A \oplus B$,  
\begin{align}
    Q =
    \begin{pmatrix}
        Q_{AA} & Q_{AB} \\
        Q_{BA} & Q_{BB}
    \end{pmatrix},
\end{align}
closure of $A$ under $Q$ requires $Q_{AB} = 0$. Since $Q$ is unitary, this implies $Q_{AA}$ is unitary. Conversely, if $Q_{AA}$ is unitary, then $A$ is invariant under $Q$. Therefore, we quantify quasisymmetry by measuring how close $Q_{AA}$ is to unitary with the metric
\begin{align}
    \epsilon = \sqrt{\frac{1}{N_A}\,\text{Tr}\!\left[P_A Q^{\dagger} P_A Q\right]},
\end{align}
where $P_A = \sum_{n\in A}\ketbra{n}{n}$ is the projection operator into subspace $A$, and $N_A$ is its dimension. The metric $\epsilon \in [0,1]$ reaches $\epsilon \equiv 1$ if and only if $A$ is exactly invariant under $Q$. When $\epsilon \approx 1$, $Q$ is a quasisymmetry, and this matrix representation identifies quasi-irreps $\Gamma_n^{(Q)}$ of $\mathcal{G}_Q$ constituting the subspace $A$.

\subsection{SOC as perturbation}
\label{sec:SOCpert}

Quasisymmetry implies that within the QS subspace, the first-order matrix element $\mel{m}{H^{\rm SOC}}{n} \approx 0$. Since quasisymmetry is not exact, this element does not vanish identically. We develop a numerical approach to calculate this matrix element using DFT and Wannierization via \texttt{wannier90}. The \kp method cannot be used for this purpose due to DFT limitations (see Appendix \ref{sec:SOC_kp}).

We generate \textit{ab initio} band structures with both scalar relativistic (SR, without SOC) and fully relativistic (FR, with SOC) pseudopotentials using QE. From these, we build tight-binding models $H_{SR}$ and $H_{FR}$. To compare $H_{SR}$ and $H_{FR}$, both models must be written in compatible basis sets. This is nontrivial, since maximal localization in \texttt{wannier90} can yield different Wannier functions for $H_{SR}$ and $H_{FR}$. To ensure consistency, we use the same initial projections for both $H_{SR}$ and $H_{FR}$, and use one of the following approaches: skip maximal localization (\texttt{num\_iter = 0}), use symmetry-adapted Wannier functions \cite{SAWF2013, WannierBerri}, or the selectively localized Wannier functions method \cite{SLWF2014}. 

\begin{figure}[b]
    \centering
    \includegraphics[width=\columnwidth]{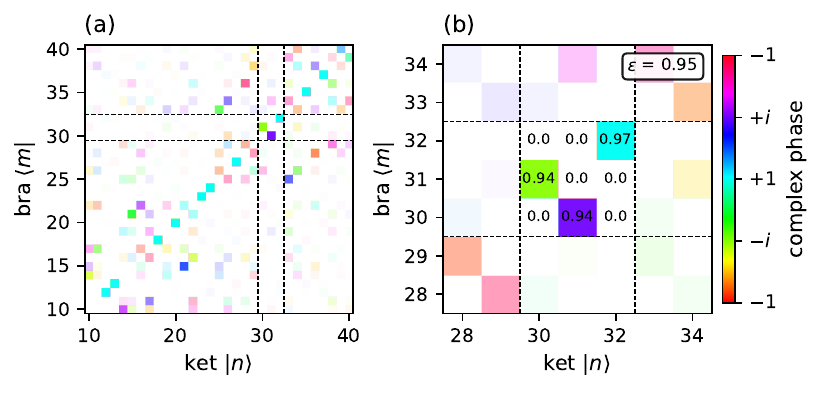}
    \caption{Complex matrix $Q_{m,n} = \mel{m}{M_y}{n}$, with the RL and ZL subspaces highlighted by the dashed lines. The values within the squares of the diagram correspond to $|Q_{m,n}|^2$, which also defines the color intensity, while the color itself represents the complex phase of the matrix element. The matrix is shown on a large scale in (a), while in (b) we enlarge the RL and ZL subspaces.}
    \label{fig: Mirror_SnSiC}
\end{figure}

Assuming the basis sets are consistent, we define
\begin{equation}
    H^{\rm SOC} \approx H_{FR} - H_{SR} \otimes \sigma_0,
\end{equation}
where $\sigma_0$ is the identity in spin space. 
The atomic from $H^{\rm SOC} \approx \sum_i \lambda_i \bm{L}_i\cdot\bm{S}$ is obtained only when Wannier functions resemble atomic orbitals, which is not guaranteed in general. Nevertheless, even when this is not the case, projecting $H^{\rm SOC}$ into the eigenstates of $H_{SR} \otimes \sigma_0$ at the k-point of interest allows us to directly compare with \kp models and identify matrix element values.

\section{Results}
\label{sec:results}

We illustrate examples of both the emergent-symmetry and inheritance pictures for quasisymmetry. For the emergent-symmetry picture we consider Sn/SiC, monolayer TMDs, and wurtzite materials. The metric $\epsilon$ is calculated for examples from the emergent-symmetry picture: Sn/SiC, TMDs and wurtzites. The intensity of first-order SOC is calculated for the TMDs only, since for Sn/SiC we could not guarantee the basis set consistency between $H_{SR}$ and $H_{FR}$ due to highly entangled bands, and for wurtzites the relevant perturbation is the crystal field, not SOC. For the inheritance picture we consider AgLa, where we identify that the first-order SOC is dominated by the $\delta\bm{k}$-term in Eq.~\eqref{eq:SecondPicture}. Throughout the paper, the single group irrep nomenclature follows Mulliken Symbols \cite{Mulliken1955}, while the double group irrep classification is based on Koster's book \cite{koster1963properties} and the Bilbao Crystallographic Server \cite{Aroyo2011Bilbao1, Aroyo2006Bilbao2, Aroyo2006Bilbao3}.

\subsection{Quasisymmetries I: Emergent symmetry picture}

\subsubsection{Sn/SiC}

We consider the Sn/SiC(0001)-$1\times1$ surface, a triangular lattice of Sn atoms adsorbed on a 6H-SiC (0001) substrate [Fig.~\ref{fig:SnSiC}(a)]. The Brillouin zone is shown in Fig.~\ref{fig:SnSiC}(b). This system hosts quasidegenerate states near the $K$ point, with conflicting interpretations in the literature \cite{Yaji2019, Tao2023}. In Ref.~\cite{Yaji2019}, the authors explain the near degeneracy as an emergent mirror $M_y$ due to the wavefunction localization along atoms aligned with the Sn sublattice. Interestingly, their explanation was developed prior to the quasisymmetry proposal \cite{Guo2022}, and it matches our emergent symmetry picture, which is further supported by the result in the Supplemental Material of Ref. \cite{PRL2024}. In contrast, Ref.~\cite{Tao2023} assumes atomistic SOC to identify an SU(2) quasisymmetry. We unify this apparent discrepancy by showing that the mirror $M_y$ quasisymmetry is consistent with the SU(2) interpretation of Ref.~\cite{Tao2023}, where $M_y$ and its $C_3$-conjugated partners act as spin rotations within the quasisymmetric subspace.

The band structure without SOC [Fig.~\ref{fig:SnSiC}(c)] shows spin-degenerate bands near the Fermi level. With SOC, a small gap of $\sim$2 meV emerges at $K$ [Fig.~\ref{fig:SnSiC}(d)], forming a Rashba-like (RL) splitting. Neighboring bands exhibit Zeeman-like (ZL) behavior with a much larger SOC gap [Fig.~\ref{fig:SnSiC}(e)]. These contrasting scales suggest quasisymmetries. The RL bands (dominated by $s$ and $p_{z}$ orbitals from Sn atoms and underlying SiC) are localized beneath the Sn sublattice. Although the little group $C_3$ at $K$ lacks mirror symmetry, this orbital localization [Figs.~\ref{fig:SnSiC}(f) and \ref{fig:SnSiC}(g)] suggests emergent mirror symmetry $M_y$ \cite{Yaji2019} within both RL and ZL subspaces.

To quantify quasisymmetry beyond visual orbital arguments, Fig.~\ref{fig: Mirror_SnSiC} shows the matrix elements $Q_{m,n} = \langle m | M_y | n\rangle$. The $Q$ matrix within both RL and ZL subspaces is nearly unitary, yielding $\epsilon = 0.968$ for the combined subspace. This confirms that $M_y$ reflects a robust constraint on the wavefunctions.

The $M_y$ quasisymmetry extends the group to $C_{3v} = C_3 \rtimes \{E, M_y\}$ within both the RL and ZL regions. The RL pair $\{\ket{R,\uparrow}, \ket{R,\downarrow}\}$ forms a two-dimensional subspace, whose orbital part transforms as the irrep $A$ of $C_3$, which becomes the quasi-irrep $A_1$ under $C_{3v}$. Under the crystal group $C_3$ the selection rules for the $z$ component of the orbital part of the SOC matrix element, $\bra{R}H_{z}\ket{R}$ [see Eq.~\eqref{eq:HsocSigma}], give us $A \otimes A \otimes A = A$, indicating it is finite. However, under the QS group $C_{3v}$ the first-order matrix element vanishes, since now we have $A_1 \otimes A_2 \otimes A_1 = A_2$. The ZL set $\{\ket{Z_+,\uparrow}, \ket{Z_+,\downarrow}, \ket{Z_-,\uparrow}, \ket{Z_-,\downarrow}\}$ also transforms under the QS group $C_{3v}$, but it does not enjoy this cancellation, since under $C_3$ the selection rules read $E \otimes A \otimes E = 2A\oplus E$, and for the QS group it reads $E\otimes A_2 \otimes E = A_1 \oplus A_2 \oplus E$, such that both cases include the trivial irrep or quasi-irrep. More generally, the effective Hamiltonian at K under the basis composed of the RL and ZL bands reads as
\begin{widetext}
    \begin{align} \label{Eq: SnSiC_H_My}
        H(K) = 
        \left(
        \begin{array}{cc|cccc}
            \epsilon_0^{R} + \delta^{R,R}_{z}& 0 & 0 & 0 & 0 & \Delta_{x}^{R,Z_-} \\
            0 & \epsilon_0^{R} -\delta^{R,R}_{z} & -\Delta_{x}^{R,Z_+} & 0 & 0 & 0 \\
            \hline
            0 & -\Delta_{x}^{R,Z_+} & \epsilon_{0}^{Z_{+}} +\Delta_{z}^{Z_+} & 0 & 0 & 0 \\
            0 & 0 & 0 & \epsilon_{0}^{Z_+} -\Delta_{z}^{Z_+} & -i \Delta_{y}^{Z_+, Z_-} & 0 \\
            0 & 0 & 0 & i \Delta_{y}^{Z_+, Z_-} & \epsilon_{0}^{Z_-} -\Delta_{z}^{ Z_-} & 0 \\
            \Delta_{x}^{R,Z_-} & 0 & 0 & 0 & 0 & \epsilon_{0}^{Z_-} +\Delta_{z}^{Z_-}
        \end{array}
        \right),
    \end{align}
\end{widetext}
where $\epsilon_0^{n}$ is the contribution from $H^0$, $\Delta_{i}^{n, m}$ and $\delta_{i}^{n, m}$ are the $\sigma_i$ SOC terms. The $\Delta_{i}^{n, m}$ terms are allowed in both the crystal group $C_3$ and the QS group $C_{3v}$, while the $\delta_{i}^{n, m}$ vanish in the QS group. The Hamiltonian above was obtained by imposing the symmetry constraints of the crystal and QS groups via the theory of invariants \cite{bir1974symmetry} with the \texttt{Qsymm} code \cite{varjas2018qsymm}. The vanishing of the $\delta_{i}^{n, m}$ terms indicates that the small SOC gap in RL bands arises, to leading order, from second-order contributions from $\Delta_{x}^{R,Z_{+}} \approx \Delta_{x}^{R,Z_{-}}$, which are constrained to be equal by the quasisymmetry, as well as the intraband couplings $\Delta_{z}^{Z_{+}} \approx \Delta_{z}^{Z_{-}}$ within the ZL sector.

To connect the $M_y$ quasisymmetry with the SU(2) symmetry proposed in Ref.~\cite{Tao2023}, notice that within the RL subspace the orbitals transform as the trivial $A_1$ quasi-irrep of $C_{3v}$ single group. Introducing the spin, the double group quasi-irrep becomes $A_1 \otimes D_{1/2} = D_{1/2}$, where $D_{1/2}$ is the spin 1/2 representation. Therefore, within this quasi-irrep the $M_y \in {\rm SU(2)}$ and its conjugated partners in $C_{3v}$, constitute a sufficient subset of SU(2) operators that defines quasisymmetry in Sn/SiC.

\begin{figure}[t]
    \centering
    \includegraphics[width=\columnwidth]{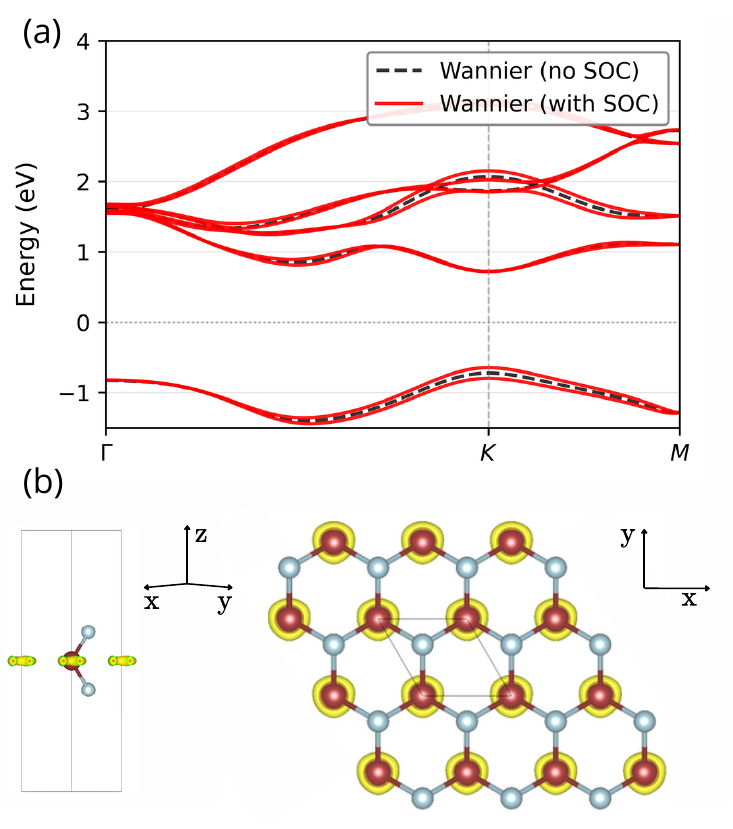}
    \caption{(a) Band structure of MoS$_2$ calculated using \texttt{Quantum ESPRESSO} and \texttt{Wannier90}, without SOC (black dashed lines) and with SOC (red solid lines). (b) Side and top views of the charge density associated with the first conduction band of MoS$_2$ at K. The brown and cyan spheres represent Mo and S atoms, respectively, while the yellow isosurfaces correspond to the real-space charge density $|\psi(\bm{r})|^2$.}
    \label{fig: Bands_TMDs}
\end{figure}

\subsubsection{Monolayer TMDs}

A similar mechanism appears in monolayer TMDs such as MoS$_2$ and WSe$_2$, which belong to space group $P\overline{6}m2$ ($D_{3h}^1$) and exhibit unusually small SOC splitting in the first conduction band at the $K$ point (little group $C_{3h}$), as shown in Fig.~\ref{fig: Bands_TMDs}. The standard explanation from Ref.~\cite{PhysRevB.88.245436} uses a tight-binding model with $H_{\text{SOC}}=\lambda \bm{L}\cdot\bm{S}$, justified by strong Wannier orbital localization. The $m=0$ character of the dominant $d_{z^{2}}$ orbital suppresses first-order SOC. This interpretation is correct within the spherical $\bm{L}\cdot\bm{S}$ approximation.

\begin{figure*}[tb] 
    \centering
    \includegraphics[width=\textwidth]{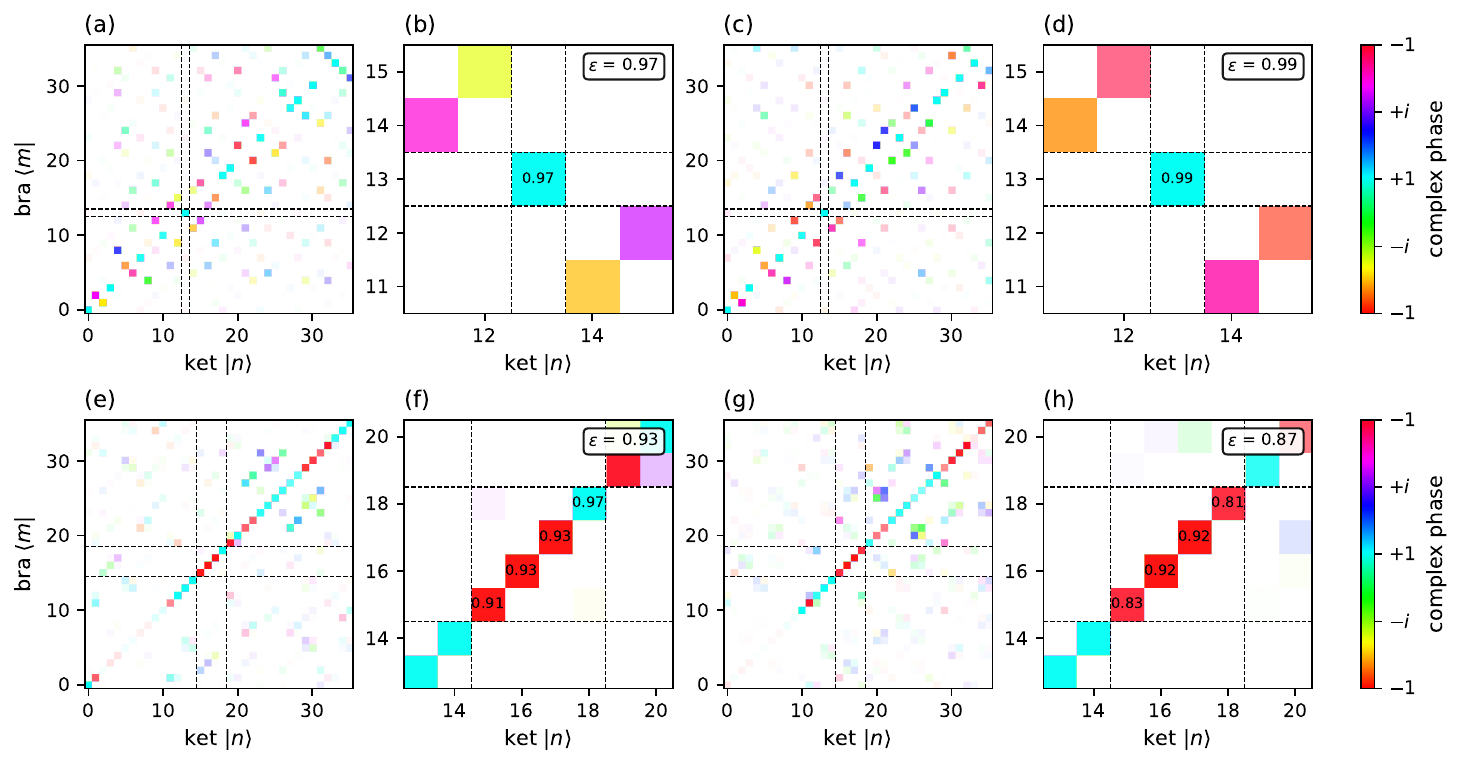}
    \caption{(a),(c) Complex matrices $Q_{m,n} = \mel{m}{\mathcal{M}_y}{n}$ for the TMD monolayers MoS$_2$ and WSe$_2$, respectively. The regions enclosed by dashed lines correspond to the QS subspaces. (b),(d) Enlargement of the QS subspaces of (a),(c). (e),(g) Equivalent matrices, $Q_{m,n} = \mel{m}{\mathcal{I}}{n}$, for the wurtzites GaN and GaP, respectively, and magnified view of the QS subspaces shown in (f),(h). The color codes and labels are equivalent to those in Fig.~\ref{fig: Mirror_SnSiC}.}
    \label{fig: Matrix_TMD_Wurtzite}
\end{figure*}

Our quasisymmetry analysis generalizes this result. As shown in Fig.~\ref{fig: Bands_TMDs}(b), the conduction-band wavefunction is strongly localized on the transition-metal sublattice, yielding an $M_y$ mirror quasisymmetry, enlarging the low-energy subspace symmetry to $D_{3h} = C_{3h} \rtimes \{E, M_y\}$. The $M_y$ matrices in Figs.~\ref{fig: Matrix_TMD_Wurtzite}(a)-\ref{fig: Matrix_TMD_Wurtzite}(d) confirm approximate unitary structure with $\epsilon > 0.95$ for both MoS$_2$ and WSe$_2$. Consequently, as in Sn/SiC, matrix elements of $H_{\text{SOC}}$ are suppressed by the full operator $\bm{\nabla}V\times\bm{p}$, not only by its atomic-limit $\bm{L}$. 

To quantify the first- and higher-order contributions to the SOC gap, we construct a Wannier representation using \texttt{wannier90}. The Wannier band structure [Fig.~\ref{fig: Bands_TMDs}(a)] matches established DFT data for MoS$_2$ \cite{Zhu2011,PhysRevB.88.245436}. The effective model at $K$ reads $H_K = \lambda_\nu \sigma_z$, where $\lambda_C$ and $\lambda_V$ are SOC intensities in conduction and valence bands. Following Sec.~\ref{sec:SOCpert}, we find that the first-order contribution, $\lambda_C^{(1)} \approx 1.78$~meV, consists of $\sim 60 \%$ of the total SOC gap $\lambda_C \approx 3.04$~meV in the conduction band, i.e., the first-order contribution is comparable to the second- and higher-order corrections, confirming that quasisymmetry suppresses the first-order term. In the valence band, which is not quasisymmetric, $\lambda_V^{(1)} \approx 153.7$~meV nearly equals the full gap $\lambda_V \approx 153.8$~meV. Similar behavior appears in WSe$_2$: For the conduction band the first-order contribution is $\lambda_C^{(1)} = 10.6$~meV, corresponding to $\sim 25\%$ out of the total $\lambda_C = 40.6$~meV, while for the valence band $\lambda_V^{(1)} = 488$~meV, and $\lambda_V = 475$~meV. 

\begin{table}[t]
    \caption{Irreps of the valence and conduction bands of wurtzite materials under the crystal  group $C_{6v}^4$, the inversion quasisymmetry group $D_{6h}^4$, and the quasicubic quasisymmetry group $T_d^2$. For the $T_d^2$ group the $[X(X^2-3Y^2)]$-like orbital ($d$-like orbital) is part of a three-dimensional irrep and its compatibility with its $C_{6v}^4$ and $D_{6h}^4$ counterparts is ill-defined. For each group the top lines refer to the single group irreps, while the bottom lines refer to the corresponding double group irreps \cite{koster1963properties, Aroyo2011Bilbao1, Aroyo2006Bilbao2, Aroyo2006Bilbao3}.}
    \label{tab:Wurtzite_irreps}
    \begin{ruledtabular}
    \begin{tabular}{c|ccc}
                   & Valence                        & Conduction GaN & Conduction GaP       \\
                   & $(X,Y,Z)$                      & $S$            & $X(X^2-3Y^2)$         \\
        \hline
        $C_{6v}^4$ & $\Gamma_1 \oplus \Gamma_5$     & $\Gamma_1$     & $\Gamma_3$    \\
                   & $2\Gamma_7\oplus \Gamma_9$     & $\Gamma_7$     & $\Gamma_8$ \\
        \hline
        $D_{6h}^4$ & $\Gamma_2^- \oplus \Gamma_6^-$ & $\Gamma_1^+$   & $B_{1u}$ \\
                   & $2\Gamma_7^-\oplus \Gamma_9^-$ & $\Gamma_7^+$   & $\Gamma_8^-$ \\
        \hline
        $T_d^2$    & $\Gamma_4$                     & $\Gamma_1$     & -  \\
                   & $\Gamma_7\oplus\Gamma_8$       & $\Gamma_6$     & -
    \end{tabular}
    \end{ruledtabular}
\end{table}

\subsubsection{Wurtzites}
\label{sec:Wurtzite}

Wurtzite materials such as GaN, GaP, and ZnO belong to space group $P6_3mc$ ($C_{6v}^4$) with little group $C_{6v}$ at $\Gamma$. Their band structures are commonly modeled using the quasicubic approximation \cite{Hopfdeld1960, Cardona1966, Voon1996, ChuangChang1996, Gutsche1967}, treating wurtzite as a strained zincblende (space group $F\bar{4}3m$ or $T_d^2$) along the $[111]$ direction. Thus, the quasicubic approximation can be interpreted as a \textit{pseudosymmetry}, assuming the wurtzite lattice is close to a cubic one. This can now be interpreted as a special case from the first quasisymmetry picture. Since $C_{6v}^4$ is not a subgroup of $T_d^2$, the quasicubic approximation does not involve direct group extension as in picture I. Instead, it operates through compatibility relations established via the common subgroup $C_3^1 \subset C_{6v}^4$ and $C_3^1 \subset T_d^2$. Additionally, we introduce a second quasisymmetry for wurtzites based on spatial inversion $\mathcal{I}$, yielding the QS group $D_{6h}^4 = C_{6v}^4 \otimes \{E, \mathcal{I}\}$, and compare its constraints with those from the quasicubic approximation. The compatible irreps for each band under the $C_{6v}^4$, $T_d^2$ and $D_{6h}^4$ groups are shown in Table~\ref{tab:Wurtzite_irreps}.

\begin{figure}[t]
    \centering
    \includegraphics[width=\columnwidth]{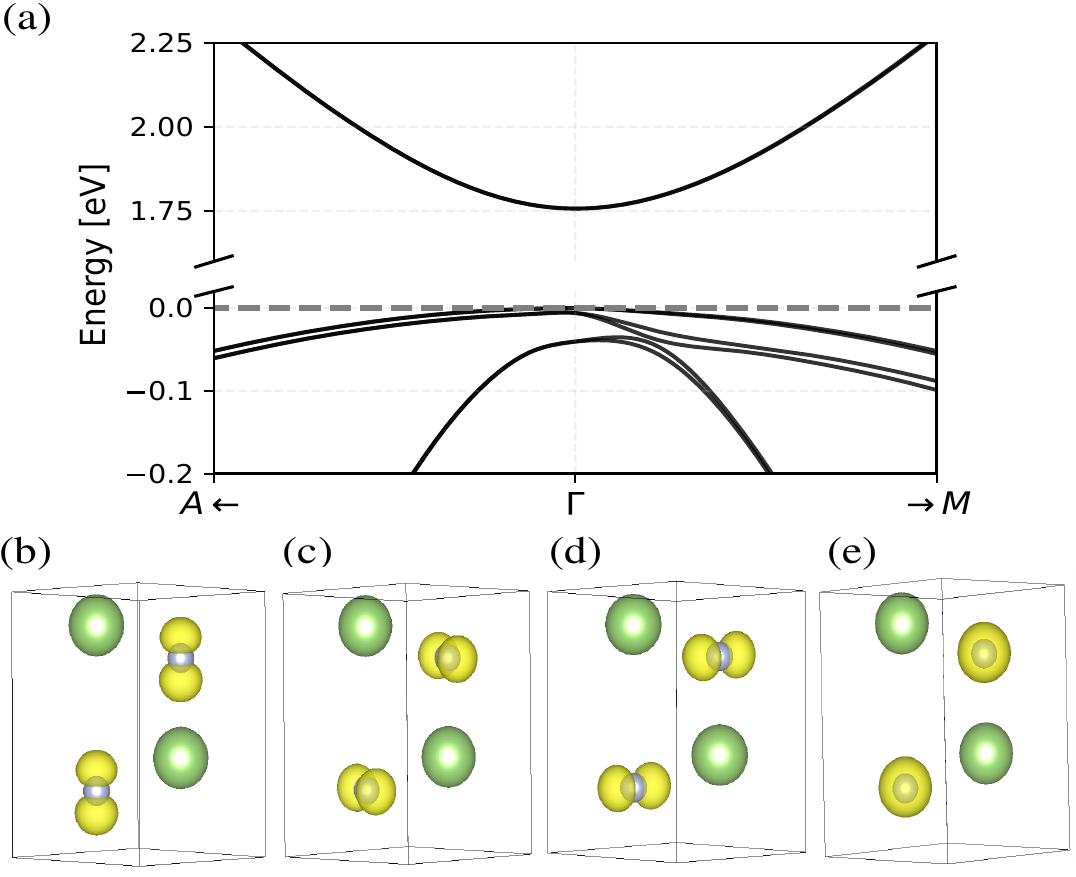}
    \caption{(a) Band structure of wurtzite GaN obtained from \texttt{Quantum ESPRESSO} along the high-symmetry path A–$\Gamma$–M. The energy axis is broken to suppress an intermediate energy range and improve the visualization of the valence and conduction bands. An enlargement in the $k$ axis around the $\Gamma$ point along A–$\Gamma$–M highlights the bands in the vicinity of $\Gamma$. (b)–(e) Charge density distributions (yellow isosurfaces) at $\Gamma$ corresponding to the three uppermost valence bands and the first conduction band. Green and gray spheres denote Ga and N atoms.}
    \label{fig: Bands_Wurtzite}
\end{figure}

A typical band structure for wurtzite crystals is shown in Fig.~\ref{fig: Bands_Wurtzite}(a) for GaN. The top valence bands are composed by $p$-like orbitals split into the $C_{6v}^4$ irreps $\Gamma_5$ for $\ket{X\pm iY}$, and $\Gamma_1$ for $\ket{Z}$ due to the hexagonal crystal field. Here, $\ket{X \pm iY}$ and $\ket{Z}$ denote $p$-orbital states combinations. The $6\times 6$ \kp Hamiltonian including SOC \cite{ChuangChang1996, willatzen2009kp} reads
\begin{equation}
    H_{\text{WZ}} =
    \begin{pmatrix}
    F & -K^* & -H^* & 0 & 0 & 0 \\
    -K & G & H & 0 & 0 & \Delta \\
    -H & H^* & \lambda & 0 & \Delta & 0 \\
    0 & 0 & 0 & F & -K & H \\
    0 & 0 & \Delta & -K^* & G & -H^* \\
    0 & \Delta & 0 & H^* & -H & \lambda
    \end{pmatrix},
    \label{eq:HWZ}
\end{equation}
where $F = E_v + \Delta_1 + \Delta_2 + \lambda + \theta$, $G = E_v + \Delta_1 - \Delta_2 + \lambda + \theta$, $\lambda = E_v + A_1 k_z^2 + A_2 (k_x^2 + k_y^2)$, $\theta = A_3 k_z^2 + A_4 (k_x^2 + k_y^2)$, $K = A_5 k_+^2$, $H = A_6 k_+ k_z + iA_7 k_+$, $\Delta = \sqrt{2}\Delta_3$, and $k_{\pm} = k_x \pm i k_y$. The parameters $\Delta_i$ represent crystal-field and SOC splittings, while $A_i$ are effective mass parameters. This model includes all symmetry-allowed couplings within the $C_{6v}^4$ group, neglecting only the k-dependent SOC (kSOC). For the full model, including kSOC, see \cite{faria2016realistic, Fu2020Wurtzite}. Particularly, the valence band energies and SOC terms are defined by the matrix elements
\begin{align}
    E_v &= \bra{Z}H^0\ket{Z},
    \label{eq: QCEv}
    \\
    E_v + \Delta_1 &= \bra{X}H^0\ket{X} = \bra{Y}H^0\ket{Y},
    \label{eq: QCD1}
    \\
    \Delta_2 &= i \bra{X}H_{z}\ket{Y},
    \label{eq: QCD2}
    \\
    \Delta_3 &= i \bra{Y}H_{x}\ket{Z} = i \bra{Z}H_{y}\ket{X},
    \label{eq: QCD3}
\end{align}
where $H_{\sigma}$ is the orbital part of the SOC operator, Eq.~\eqref{eq:HsocSigma}.

In the quasicubic approximation, one considers $\{\ket{X}, \ket{Y}, \ket{Z}\}$ approximately as partners under the $\Gamma_4$ irrep of $T_d^2$. This yields $\Delta_2 = \Delta_3$, and $\Delta_1 \approx 0$, but a correction $\Delta_1 \approx \Delta_{\rm cr}$ is kept to account for the hexagonal crystal field. If we consider that the wurtzite potential can be written as $V_{\rm WZ} = V_{\rm ZB} + \Delta V$, where $V_{\rm ZB}$ is the zincblende potential and $\Delta V$ is the perturbation that breaks the cubic symmetry, then $\Delta_1 = \bra{Z}\Delta V\ket{Z} - \bra{X}\Delta V\ket{X} \approx 0$ to first-order in Löwdin perturbation theory, since $\ket{X}$ and $\ket{Z}$ are partners under the cubic group. Consequently, $\Delta_1 \approx \Delta_{\rm cr}$ arises from second-order contributions, analogous to the emergent quasisymmetry picture from Sec.~\ref{sec:emergent}, but achieved through compatibility relations rather than group extension. Moreover, to leading order one obtains $A_1 - A_2 = -A_3 = 2A_4$, $A_3 + 4A_5 = \sqrt{2}A_6$, and $A_7 = 0$ \cite{ChuangChang1996}. The quasicubic validity is typically verified by $\Delta_3/\Delta_2 \approx 1$. From DFT-fitted parameters [Fig.~\ref{fig: Bands_Wurtzite}(a)], we find that $\Delta_3/\Delta_2 \sim 1$ for GaP and CdS, while it deviates from the perfect ratio by 20\% -- 50\% for GaN and ZnO \cite{Langer1970, Suzuki1995, Yamaguchi1998, Dugdale2000wurtzite, Lambrecht2002}. Particularly, for GaP the first and second conduction bands are inverted \cite{Assali2016, daSilva2020}, such that the first conduction band has $X(X^2-3Y^2)$ character, which does not have a well-defined partner under the cubic group, and thus its compatibility with the $C_{6v}^4$ irreps is ill-defined (see Table \ref{tab:Wurtzite_irreps}), breaking the quasicubic approximation.

In contrast, inversion $\mathcal{I}$ acts as an alternative quasisymmetry within the subspace of top valence bands and lowest conduction bands, yielding quasisymmetry group $D_{6h}^4 = C_{6v}^4 \otimes \{E, \mathcal{I}\}$. Wavefunctions in this subspace are strongly localized on anion sites [Figs.~\ref{fig: Bands_Wurtzite}(b)-\ref{fig: Bands_Wurtzite}(e)]. The representation matrices $Q_{mn} = \bra{m}\mathcal{I}\ket{n}$ [Figs.~\ref{fig: Matrix_TMD_Wurtzite}(e)-\ref{fig: Matrix_TMD_Wurtzite}(h)] yield $\epsilon \sim 0.9$ for GaN, GaP and CdS, while for ZnO we get $\epsilon \sim 0.76$, confirming approximate inversion symmetry. Diagonal entries of $Q_{mn}$ show valence bands are odd under inversion (p-like), while conduction bands are even (s-like) for GaN, CdS and ZnO, and odd for GaP \cite{Assali2016, daSilva2020}. Accordingly, these bands can be described by selection rules involving inversion symmetry, which constrain first-order perturbation terms within the subspace, as well as matrix elements of relevant operators. The constraints induced by $\mathcal{I}$ on the Hamiltonian from Eq.~\eqref{eq:HWZ} are less restrictive than those of the quasicubic approximation; they lead only to $A_7 = 0$, $\alpha_i = 0$ (kSOC terms), which is reasonable for all wurtzites. 

The inversion quasisymmetry constraints can be further applied to optical selection rules involving the first conduction band of GaP, which is odd under quasi-inversion in the QS group $D_{6h}^4$. The optical transition oscillator strengths depend on electric dipole matrix elements $\bm{d} = -e \bm{r}$. Within the crystal double group $C_{6v}^4$, the $\Gamma_9$ valence to $\Gamma_8$ conduction band transition in GaP is allowed for light perpendicular to the $c$ axis, since $\Gamma_9 \otimes \Gamma_8 = \Gamma_5(x,y) \oplus \Gamma_6$. Under quasisymmetry group $D_{6h}^4$, the transition becomes first-order forbidden because both bands belong to odd quasi-irreps; thus $\Gamma_9^- \otimes \Gamma_8^- = \Gamma_5^+ \oplus \Gamma_6^+$ contains no vector irrep. Moreover, the suppression of this matrix element due to QS can be quantified via the metric $\epsilon$. We can split the wavefunctions into $\ket{\Gamma_\nu} = \sqrt{\epsilon}\ket{\Gamma_\nu^-} + \sqrt{1-\epsilon}\ket{\Gamma_\nu^+}$, where $\ket{\Gamma_\nu^\mp}$ carry the odd (quasisymmetric) and even (small correction) components. Thus, the dipole matrix element can be written as 
\begin{multline}
    \bra{\Gamma_9}\bm{d}\ket{\Gamma_8} = 
    \epsilon \bra{\Gamma_9^-}\bm{d}\ket{\Gamma_8^-} + 
    (1-\epsilon)\bra{\Gamma_9^+}\bm{d}\ket{\Gamma_8^+}
    \\
    \sqrt{\epsilon(1-\epsilon)}\big[
        \bra{\Gamma_9^-}\bm{d}\ket{\Gamma_8^+} + 
        \bra{\Gamma_9^+}\bm{d}\ket{\Gamma_8^-} 
    \big].
\end{multline}
The first and second terms vanish by symmetry constraints, while the others terms are suppressed, since the spectral weight of the $\ket{\Gamma_\nu^+}$ component is diminished outside the QS subspace, as shown in Fig.~\ref{fig: Matrix_TMD_Wurtzite}. For GaP, with $\epsilon \sim 0.9$, the factor $\sqrt{\epsilon(1-\epsilon)} \sim 0.3$ indicates a $\sim 70\%$ suppression of the transition intensity compared to a fully allowed transition, consistent with experimental observations of weak A-transitions in GaP \cite{Assali2016}.

\begin{figure*}[t]
    \centering
    \includegraphics[width=\textwidth]{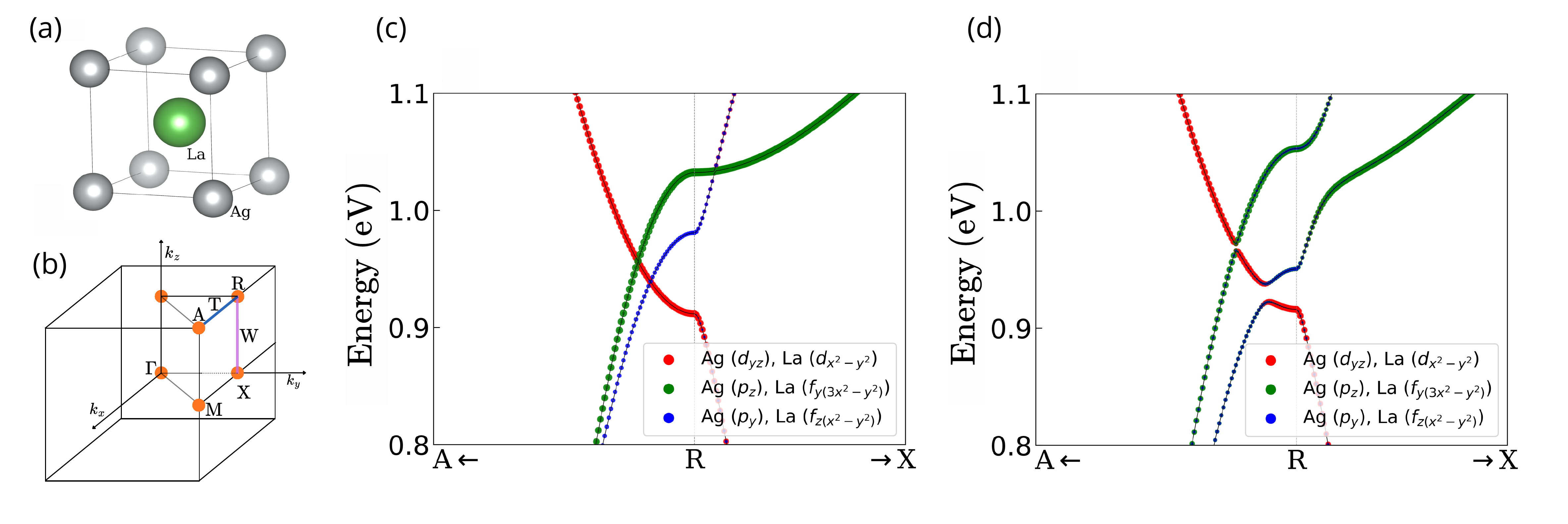}
    \caption{(a) Simple cubic unit cell of AgLa, with La (green) and Ag (gray) atoms. (b) Brillouin zone of the AgLa crystal, highlighting the high-symmetry paths T = R--A and W = R--X. (c),(d) AgLa band structure along the A--R--X path without and with spin-orbit coupling, respectively. The color coding denotes the dominant orbital-projected contributions indicated in the legend. SOC lifts the band degeneracies at the crossing points, inducing band splittings. The splittings are noticeably smaller along the T path, in contrast with the larger splitting along the W path.}
    \label{fig: Painel_AgLa}
\end{figure*}

\subsection{Quasisymmetries II: Inheritance picture}
\label{sec:AgLa}

We revisit AgLa \cite{PRL2024} to clarify its quasisymmetry.
AgLa [Fig.~\ref{fig: Painel_AgLa}(a)] is a tetragonal compound in space group $P4/mmm$ ($D_{4h}^1$).
The Brillouin zone is shown in Fig.~\ref{fig: Painel_AgLa}(b). Without SOC, the band structure exhibits crossings along paths A--R (denoted T) and R--X (denoted W) involving three states $\ket{\alpha}$, $\ket{\beta}$, and $\ket{\gamma}$, dominated respectively by orbitals 
$\alpha = {\rm Ag}(d_{yz}) + {\rm La}(d_{x^{2}-y^{2}})$,
$\beta = {\rm Ag}(p_{z}) + {\rm La}(f_{y(3x^{2}-y^{2})})$, and 
$\gamma = {\rm Ag}(p_{y}) + {\rm La}(f_{z(x^{2}-y^{2})})$. Including SOC, Figs.~\ref{fig: Painel_AgLa}(c) and \ref{fig: Painel_AgLa}(d) show anticrossing gaps are small along the T-path ($\sim$7 meV for $\alpha-\beta$, $\sim$20 meV for $\alpha-\gamma$ anticrossings), while along the W-path the $\beta-\gamma$ anticrossing gap is larger ($\sim$80 meV).

Both T and W paths are invariant under $C_{2v}$, but with different axes: $C_{2v}^T = \{E, C_2(x), M_z, M_y\}$ and $C_{2v}^W = \{E, C_2(z), M_y, M_x\}$. Table~\ref{tab:irrepsAgLa} shows irreps of $\alpha$, $\beta$, $\gamma$ under these groups and $D_{2h}$ at R. Selection rules along T give $A_1 \otimes B_2 = B_2$ and $A_1 \otimes B_1 = B_1$ for $\alpha-\beta$ and $\alpha-\gamma$ crossings, indicating finite SOC from $\sigma_z$ and $\sigma_y$ couplings. Along W, the $\beta-\gamma$ crossing requires $\sigma_x$ since $B_2 \otimes A_1 = B_2$. Therefore, under crystal groups $C_{2v}^T$ and $C_{2v}^W$, all crossings in Fig.~\ref{fig: Painel_AgLa}(c) should open SOC gaps. 

\begin{table}[t]
    \caption{Irreducible representations for the $\alpha$, $\beta$, and $\gamma$ states, and the angular momentum $R_\nu$ (orbital part of the SOC operators) for $\nu=\{x,y,z\}$ under the symmetry groups discussed in Sec.~\ref{sec:AgLa}.}
    \label{tab:irrepsAgLa}
    \begin{ruledtabular}
    \begin{tabular}{c|ccc|ccc}
                   & $\alpha$ & $\beta$  & $\gamma$ & $R_x$    & $R_y$    & $R_z$ \\
        \hline
        $C_{2v}^T$ & $A_1$    & $B_2$    & $B_1$    & $A_2$    & $B_1$    & $B_2$ \\
        $C_{2v}^W$ & $A_1$    & $B_2$    & $A_1$    & $B_2$    & $B_1$    & $A_2$ \\
        $D_{2h}^R$ & $A_g$    & $B_{2u}$ & $B_{1u}$ & $B_{3g}$ & $B_{2g}$ & $B_{1g}$ \\
    \end{tabular}
    \end{ruledtabular}
\end{table}

Reference~\cite{PRL2024} shows quasisymmetries can emerge when $C_{2v}$ is extended by $S_{4}$, $C_{4}$, or inversion $\mathcal{I}$ symmetries. For crossings in Fig.~\ref{fig: Painel_AgLa}(c), these extensions do not leave T or W paths invariant, ruling out the emergent symmetry picture from Sec.~\ref{sec:emergent} and Ref.~\cite{PRL2024} itself. Instead, spatial inversion $\mathcal{I}$ acts as a QS under the inheritance picture (Sec.~\ref{sec:kppicture}), yielding $D_{2h}=C_{2v}\otimes\{E, \mathcal{I}\}$ as the quasisymmetry group from proximity to the R point, without requiring T or W path invariance.

To illustrate the QS constraints, we analyze SOC matrix elements for each crossing. From Eq.~\eqref{eq:SecondPicture} and selection rules from Table~\ref{tab:irrepsAgLa}, the symmetry-allowed contributions are
\begin{align}
    \bra{\alpha} H_{z} \ket{\beta}_T &= -\frac{1}{4}\alpha^2\bra{\alpha} V_y \ket{\beta}_R \delta k_x,
    \\
    \bra{\alpha} H_{y} \ket{\gamma}_T &= +\frac{1}{4}\alpha^2\bra{\alpha} V_z \ket{\gamma}_R \delta k_x,
    \\
    \bra{\beta} H_{x} \ket{\gamma}_W &= +\frac{1}{4}\alpha^2\bra{\beta} (V_y p_z - V_z p_y) \ket{\gamma}_R,
\end{align}
where $H_\sigma$ is defined in Eq.~\eqref{eq:HsocSigma}, $V_\nu = \partial V/ \partial x_\nu$. The indices on the bra-kets indicate that the matrix elements are evaluated at the crossings along the T or W paths, or at the R point. For the crossings along T, only $\delta k$-terms are allowed, while along W, only p-terms contribute. Since $\delta k$-SOC terms are typically small in \kp models, k-term contributions are smaller than p-terms, explaining the smaller gaps along T versus W.

\section{Conclusions}
\label{sec:conclusions}

We established a comprehensive framework for characterizing quasisymmetry (QS) in crystalline materials combining group-theoretical analysis, \kp modeling, and first-principles simulations. We consider quasisymmetries emerging through two distinct and nonequivalent pictures: (i) wavefunction localization on sublattices invariant under the quasisymmetry, or (ii) quasisymmetry inheritance from nearby high-symmetry points. Our approach extends the original quasisymmetry proposal \cite{Guo2022, Hu2023Hierarchy} and the point group classification from Ref.~\cite{PRL2024} by identifying emergence mechanisms and distinguishing between sublattice localization and inheritance pictures. This introduces \textit{predictive power} to quasisymmetries, which can be used to understand and engineer spin splittings in semiconductors, modulate optical transition strengths, and design topological phase transitions controlled by perturbations. 

The calculated representation matrices $Q_{m,n}$ allow us to quantify quasisymmetry via the metric $\epsilon$. Combined with the group-subgroup diagram from Ref.~\cite{PRL2024} (Fig.~2), this enables future high-throughput identification and classification of materials by quasisymmetry degree. We propose a numerical procedure to quantitatively verify first-order SOC gap intensity, confirming significant suppression of first-order contributions.

We applied our methodology to various materials. For Sn/SiC, we unified apparently contradictory interpretations: Reference \cite{Yaji2019} invokes emergent $M_y$ quasisymmetry within the RL subspace, while Ref.~\cite{Tao2023} describes an SU(2) quasisymmetry. We show that the $M_y$ quasisymmetry, and its $C_3$ conjugated partners, yield a sufficient subset of SU(2) operators within the RL subspace. For AgLa, Ref.~\cite{PRL2024} introduces inversion as quasisymmetry despite noninvariant Bloch phases along T and W paths. Our inheritance picture solves this issue by transferring matrix elements to the inversion-invariant high-symmetry point R. For wurtzites, we argue that the quasicubic approximation can be seen as a quasisymmetry generated through compatibility relations between crystal groups linked by the common subgroup $C_3^1$. Additionally, wurtzites allow for a second inversion quasisymmetry, which explains the weak A-transition oscillator strength in GaP.

\begin{acknowledgments}
    We acknowledge useful conversations with Jinwoong Kim and David Vanderbilt about approaches to calculate the spin-orbit couplings using \texttt{wannier90}, and useful conversations with Paulo E. Faria Junior about the DFT limitations to extract the spin-orbit coupling via the \kp method. This work was supported by the funding agencies CNPq, CAPES, and FAPEMIG. Particularly, E.V.C.L.~and T.M.S.~acknowledge CENAPAD-SP (Project No. proj483) and LNCC (project SCAFMat2) for the computational time, and G.J.F.~acknowledges funding from the FAPEMIG Grant No. PPM-00798-18.
\end{acknowledgments}

\section*{Data Availability}

The data supporting the findings of this study are openly available in the Zenodo repository at \cite{Zenodo}.

\appendix

\section{Computational Methodology}
\label{sec:computational}

First-principles simulations were performed within the framework of density functional theory (DFT), using the Vienna \textit{ab initio} simulation package \texttt{VASP} \cite{Kresse1993, Kresse1996} and \texttt{Quantum ESPRESSO} (QE) \cite{Giannozzi2009, Giannozzi2017}. The exchange-correlation energy was treated within the generalized gradient approximation (GGA) using the Perdew-Burke-Ernzerhof (PBE) functional~\cite{Perdew1996}. For the Sn/SiC(0001) - $1\times 1$ system, a kinetic energy cutoff of 500 eV was employed, with the Brillouin zone sampled using an 11$\times$11$\times$1 Monkhorst-Pack $k$-mesh. For the AgLa crystal, a cutoff energy of 400 eV and a 9$\times$9$\times$9 $k$-mesh were used. In the case of the wurtzite structures GaN and GaP, a higher cutoff of 800 eV was adopted alongside an 8$\times$8$\times$8 $k$-mesh to ensure convergence. For the monolayer TMD systems MoS$_2$ and WSe$_2$, all band-structure and $\bm{k} \cdot \bm{p}$ matrix-element calculations were performed using PAW pseudopotentials within QE. For MoS$_2$, we used a kinetic-energy cutoff of 100 Ry, an \texttt{ecutrho} of 800 Ry and a $10\times 10\times 1$ Monkhorst-Pack sampling. For WSe$_2$, the calculations employed a kinetic-energy cutoff of 100 Ry, an \texttt{ecutrho} of 1200 Ry, and the same $10\times 10\times 1$ grid. All results for these two systems, except for the charge-density plots, were obtained using PAW pseudopotentials. The charge densities were recalculated using norm-conserving PBE pseudopotentials, since the PAW densities produced nonphysical negative regions in the interstitial space. In all calculations, the atomic positions were fully relaxed until the total energy difference between successive steps was below $10^{-8}$ eV and the residual forces were smaller than 0.001 eV/\AA.

Moreover, throughout this paper we have used several codes to perform the different steps of our analysis. The \texttt{DFT2kp} code \cite{Cassiano2024} and the \texttt{IrRep} code \cite{IrRep2022} were used to calculate representation matrices and irreducible representations directly from DFT data. The \texttt{wannier90} \cite{mostofi2014updated, pizzi2020wannier90, mostofi2008wannier90} code  was used to construct Wannier functions and tight-binding models, while the \texttt{PythTB} \cite{PythTB} code was used to manipulate the tight-binding models. Finally, the \texttt{Qsymm} code \cite{varjas2018qsymm} was used to analyze symmetry constraints on the Hamiltonian via the theory of invariants.

All scripts and input files used here are available upon reasonable request.

\section{DFT limitations to extract SOC via \texorpdfstring{\kp}{kp}}
\label{sec:SOC_kp}

In principle, one can express the matrix elements of SOC perturbatively via the \kp method using the following approach. Consider $\ket{n}$ as spinless eigenstates of the scalar relativistic Hamiltonian $H_{SR}$, and $\ket{n,\sigma_n} = \ket{n} \otimes \ket{\sigma}$ its extension to include spin $\sigma = \{\uparrow,\downarrow\}$. The first-order matrix element of SOC then reads
\begin{equation}
    \bra{m,\sigma_m}H^{\rm SOC}\ket{n,\sigma_n} = \dfrac{\alpha^2}{4} 
    \bra{m}\nabla V \times \bm{p} \ket{n} 
    \cdot \bm{\sigma}_{m,n},
    \label{eq:matelSOC}
\end{equation}
where $\bm{\sigma}_{m,n} = \bra{\sigma_m} \bm{\sigma} \ket{\sigma_n}$. Now we focus on the orbital term, $\bra{m}\nabla V \times \bm{p} \ket{n}$ that determines the intensity of this matrix element. We can eliminate $V$ using the commutator $\nabla V = -i [V, \bm{p}] \approx -i [H^0, \bm{p}]$, often used by Cardona and collaborators \cite{Voon1996}, with $H^0$ from Eq.~\eqref{eq:H0}. The expression is considered an approximation because it neglects scalar fine structure corrections that would contribute with higher-order ($\propto \alpha^4$) terms to the matrix element of Eq.~\eqref{eq:matelSOC}. Further introducing the completeness relation, the matrix element for the $z$ component reads
\begin{equation}
    \bra{m}\nabla V \times \bm{p}\ket{n}_z = i \sum_\ell E_\ell^0 \Big[
    p_x^{(m,\ell)} p_y ^{(\ell,n)}
    - 
    (x\leftrightarrow y)
    \Big],
\end{equation}
and cyclic permutations of $(x,y,z)$ for the other components. Here $p_\nu^{(m,\ell)} = \bra{m}p_\nu\ket{\ell}$ is the matrix element of momentum that can be directly extracted from DFT data using the \texttt{DFT2kp} code \cite{Cassiano2024} together with band edge energies $E^0_\ell$. Therefore, in principle we have all elements to compute $\bra{m}\nabla V \times \bm{p}\ket{n}$ numerically. Unfortunately, this approach fails because the sum over $\ell$ must run over all states, while typical DFT output for codes like \texttt{QE} \cite{Giannozzi2009, Giannozzi2017}, \texttt{VASP} \cite{Kresse1993, Kresse1996}, and \texttt{Wien2k} \cite{WIEN2k2020} does not provide direct access to core states and in practice the sum over $\ell$ above is effectively truncated. Since core states are the ones that contribute the most to the SOC intensity, the resulting matrix element with the truncated $\ell$-sum lacks the dominant contributions. To overcome this limitation, further work is needed to incorporate momentum matrix elements including core states into \texttt{DFT2kp}, but this is beyond the scope of the current work. In contrast, the Wannierization method provides atomisticlike tight-binding models that are appropriate for estimating SOC as discussed in Sec.~\ref{sec:SOCpert}.

\bibliography{main}

@dataset{Zenodo,
  author       = {Nunes Assunção, Bryan Douglas and
                  Victor Caires Lopes, Emmanuel and
                  Schmidt, Tome M. and
                  J. Ferreira, Gerson},
  title        = {{Physical Pictures for Quasisymmetry in Crystals [Data Set]}},
  year         = 2026,
  note    = {{Zenodo}},
  doi          = {10.5281/zenodo.21476577},
  url          = {https://doi.org/10.5281/zenodo.21476577}
}

@article{Acosta2021SOC,
  title = {{Different shapes of spin textures as a journey through the Brillouin zone}},
  author = {Mera Acosta, Carlos and Yuan, Linding and Dalpian, Gustavo M. and Zunger, Alex},
  journal = {Phys. Rev. B},
  volume = {104},
  issue = {10},
  pages = {104408},
  numpages = {16},
  year = {2021},
  month = {Sep},
  publisher = {American Physical Society},
  doi = {10.1103/PhysRevB.104.104408},
  url = {https://link.aps.org/doi/10.1103/PhysRevB.104.104408}
}

@article{Wang2012TMDReview,
  title = {{Electronics and optoelectronics of two-dimensional transition metal dichalcogenides}},
  volume = {7},
  ISSN = {1748-3395},
  url = {http://dx.doi.org/10.1038/nnano.2012.193},
  DOI = {10.1038/nnano.2012.193},
  number = {11},
  journal = {Nat. Nanotechnol.},
  publisher = {Springer Science and Business Media LLC},
  author = {Wang,  Qing Hua and Kalantar-Zadeh,  Kourosh and Kis,  Andras and Coleman,  Jonathan N. and Strano,  Michael S.},
  year = {2012},
  month = nov,
  pages = {699}
}

@article{Chhowalla2013TMDReview,
  title = {{The chemistry of two-dimensional layered transition metal dichalcogenide nanosheets}},
  volume = {5},
  ISSN = {1755-4349},
  url = {http://dx.doi.org/10.1038/nchem.1589},
  DOI = {10.1038/nchem.1589},
  number = {4},
  journal = {Nat. Chem.},
  publisher = {Springer Science and Business Media LLC},
  author = {Chhowalla,  Manish and Shin,  Hyeon Suk and Eda,  Goki and Li,  Lain-Jong and Loh,  Kian Ping and Zhang,  Hua},
  year = {2013},
  month = mar,
  pages = {263}
}

@article{MakShan2016TMDReview,
  title = {{Photonics and optoelectronics of 2D semiconductor transition metal dichalcogenides}},
  volume = {10},
  ISSN = {1749-4893},
  url = {http://dx.doi.org/10.1038/nphoton.2015.282},
  DOI = {10.1038/nphoton.2015.282},
  number = {4},
  journal = {Nat. Photon.},
  publisher = {Springer Science and Business Media LLC},
  author = {Mak,  Kin Fai and Shan,  Jie},
  year = {2016},
  month = mar,
  pages = {216}
}

@article{Manzeli2017TMDReview,
  title = {{2D transition metal dichalcogenides}},
  volume = {2},
  pages = {17033},
  ISSN = {2058-8437},
  url = {http://dx.doi.org/10.1038/natrevmats.2017.33},
  DOI = {10.1038/natrevmats.2017.33},
  number = {8},
  journal = {Nat. Rev. Mater.},
  publisher = {Springer Science and Business Media LLC},
  author = {Manzeli,  Sajedeh and Ovchinnikov,  Dmitry and Pasquier,  Diego and Yazyev,  Oleg V. and Kis,  Andras},
  year = {2017},
  month = June 
}

@Article{Zhang2020JanusReview,
  author    = {Zhang, Lei and Yang, Zhenjingfeng and Gong, Tian and Pan, Ruikun and Wang, Huide and Guo, Zhinan and Zhang, Han and Fu, Xiao},
  title     = {{Recent advances in emerging Janus two-dimensional materials: from fundamental physics to device applications}},
  doi       = {10.1039/d0ta01999b},
  issn      = {2050-7496},
  number    = {18},
  pages     = {8813},
  url       = {http://dx.doi.org/10.1039/D0TA01999B},
  volume    = {8},
  fjournal  = {Journal of Materials Chemistry A},
  journal   = {J. Mater. Chem.A},
  publisher = {Royal Society of Chemistry (RSC)},
  year      = {2020},
}

@article{Han2018TMDReview,
  title = {{van der Waals Metallic Transition Metal Dichalcogenides}},
  volume = {118},
  ISSN = {1520-6890},
  url = {http://dx.doi.org/10.1021/acs.chemrev.7b00618},
  DOI = {10.1021/acs.chemrev.7b00618},
  number = {13},
  journal = {Chem. Rev.},
  publisher = {American Chemical Society (ACS)},
  author = {Han,  Gang Hee and Duong,  Dinh Loc and Keum,  Dong Hoon and Yun,  Seok Joon and Lee,  Young Hee},
  year = {2018},
  month = jun,
  pages = {6297}
}

@article{Yin2021Janusreview,
  title = {{Recent advances in low-dimensional Janus materials: theoretical and simulation perspectives}},
  volume = {2},
  ISSN = {2633-5409},
  url = {http://dx.doi.org/10.1039/D1MA00660F},
  DOI = {10.1039/d1ma00660f},
  number = {23},
  journal = {Mater. Adv.},
  publisher = {Royal Society of Chemistry (RSC)},
  author = {Yin,  Wen-Jin and Tan,  Hua-Jian and Ding,  Pei-Jia and Wen,  Bo and Li,  Xi-Bo and Teobaldi,  Gilberto and Liu,  Li-Min},
  year = {2021},
  pages = {7543}
}

@Article{Naguib2013MXenesreview,
  author    = {Naguib, Michael and Mochalin, Vadym N. and Barsoum, Michel W. and Gogotsi, Yury},
  title     = {{25th Anniversary Article: MXenes: A New Family of Two‐Dimensional Materials}},
  doi       = {10.1002/adma.201304138},
  issn      = {1521-4095},
  number    = {7},
  pages     = {992},
  url       = {http://dx.doi.org/10.1002/adma.201304138},
  volume    = {26},
  fjournal  = {Advanced Materials},
  journal   = {Adv. Mater.},
  month     = dec,
  publisher = {Wiley},
  year      = {2013},
}

@Article{Khazaei2017MXenesreview,
  author    = {Khazaei, Mohammad and Ranjbar, Ahmad and Arai, Masao and Sasaki, Taizo and Yunoki, Seiji},
  title     = {{Electronic properties and applications of MXenes: a theoretical review}},
  doi       = {10.1039/c7tc00140a},
  issn      = {2050-7534},
  number    = {10},
  pages     = {2488},
  url       = {http://dx.doi.org/10.1039/C7TC00140A},
  volume    = {5},
  fjournal  = {Journal of Materials Chemistry C},
  journal   = {J. Mater. Chem. C},
  publisher = {Royal Society of Chemistry (RSC)},
  year      = {2017},
}

@article{Guha2022MXenes,
  title = {{High-throughput design of functional-engineered MXene transistors with low-resistive contacts}},
  volume = {8},
  pages = {202},
  ISSN = {2057-3960},
  url = {http://dx.doi.org/10.1038/s41524-022-00885-6},
  DOI = {10.1038/s41524-022-00885-6},
  number = {1},
  journal = {npj Computational Materials},
  publisher = {Springer Science and Business Media LLC},
  author = {Guha,  Sirsha and Kabiraj,  Arnab and Mahapatra,  Santanu},
  year = {2022},
  month = sep 
}

@article{Zheng2024janus,
  title = {{Symmetry Manipulation of Two-Dimensional Semiconductors by Janus Structure}},
  volume = {6},
  ISSN = {2643-6728},
  url = {http://dx.doi.org/10.1021/accountsmr.4c00236},
  DOI = {10.1021/accountsmr.4c00236},
  number = {2},
  journal = {Acc. Mater. Res.},
  publisher = {American Chemical Society (ACS)},
  author = {Zheng,  Xueqiu and Zhou,  Yi and Guo,  Yunfan},
  year = {2024},
  month = nov,
  pages = {124}
}

@article{Yagmurcukardes2020janus,
  title = {{Quantum properties and applications of 2D Janus crystals and their superlattices}},
  volume = {7},
  pages = {011311},
  ISSN = {1931-9401},
  url = {http://dx.doi.org/10.1063/1.5135306},
  DOI = {10.1063/1.5135306},
  number = {1},
  journal = {Appl. Phys. Rev.},
  publisher = {AIP Publishing},
  author = {Yagmurcukardes,  M. and Qin,  Y. and Ozen,  S. and Sayyad,  M. and Peeters,  F. M. and Tongay,  S. and Sahin,  H.},
  year = {2020},
  month = feb 
}

@article{mak2010atomically,
  title = {{Atomically Thin ${\mathrm{MoS}}_{2}$: A New Direct-Gap Semiconductor}},
  author = {Mak, Kin Fai and Lee, Changgu and Hone, James and Shan, Jie and Heinz, Tony F.},
  journal = {Phys. Rev. Lett.},
  volume = {105},
  issue = {13},
  pages = {136805},
  numpages = {4},
  year = {2010},
  month = {Sep},
  publisher = {American Physical Society},
  doi = {10.1103/PhysRevLett.105.136805},
  url = {https://link.aps.org/doi/10.1103/PhysRevLett.105.136805}
}

@article{mccreary2016synthesis,
  title = {{Synthesis of Large-Area WS$_2$ monolayers with Exceptional Photoluminescence}},
  volume = {6},
  ISSN = {2045-2322},
  url = {http://dx.doi.org/10.1038/srep19159},
  DOI = {10.1038/srep19159},
  number = {1},
  journal = {Sci. Rep.},
  pages = {19159},
  publisher = {Springer Science and Business Media LLC},
  author = {McCreary,  Kathleen M. and Hanbicki,  Aubrey T. and Jernigan,  Glenn G. and Culbertson,  James C. and Jonker,  Berend T.},
  year = {2016},
  month = Jan 
}

@article{singh2020tunable,
  title = {{Tunable light emission from chemical vapor deposited two-dimensional MoSe$_2$ by layer variation and S incorporation}},
  volume = {38},
  ISSN = {1520-8559},
  url = {http://dx.doi.org/10.1116/1.5124998},
  DOI = {10.1116/1.5124998},
  number = {2},
  journal = {J. Vac. Sci. Technol.},
  publisher = {American Vacuum Society},
  author = {Singh,  Vineeta and Late,  Dattatray J. and Rath,  Shyama},
  year = {2020},
  pages = {023402}
}

@article{jones2013optical,
  title = {{Optical generation of excitonic valley coherence in monolayer WSe$_2$}},
  volume = {8},
  ISSN = {1748-3395},
  url = {http://dx.doi.org/10.1038/nnano.2013.151},
  DOI = {10.1038/nnano.2013.151},
  number = {9},
  journal = {Nat. Nanotechnol.},
  publisher = {Springer Science and Business Media LLC},
  author = {Jones,  Aaron M. and Yu,  Hongyi and Ghimire,  Nirmal J. and Wu,  Sanfeng and Aivazian,  Grant and Ross,  Jason S. and Zhao,  Bo and Yan,  Jiaqiang and Mandrus,  David G. and Xiao,  Di and Yao,  Wang and Xu,  Xiaodong},
  year = {2013},
  month = Aug,
  pages = {634}
}

@article{Testelin2022gfactor,
  title = {Land\'e $g$ factors in tetragonal halide perovskite: A multiband k\ifmmode\cdot\else\textperiodcentered\fi{}p model},
  author = {Garcia-Arellano, G. and Boujdaria, K. and Chamarro, M. and Testelin, C.},
  journal = {Phys. Rev. B},
  volume = {106},
  issue = {16},
  pages = {165201},
  numpages = {10},
  year = {2022},
  month = {Oct},
  publisher = {American Physical Society},
  doi = {10.1103/PhysRevB.106.165201},
  url = {https://link.aps.org/doi/10.1103/PhysRevB.106.165201}
}

@article{StranksSnaith2015PerovskiteReview,
  title = {{Metal-halide perovskites for photovoltaic and light-emitting devices}},
  author = {Stranks, Samuel D. and Snaith, Henry J.},
  journal = {Nat. Nanotechnol.},
  volume = {10},
  number = {5},
  pages = {391--402},
  year = {2015},
  doi = {10.1038/nnano.2015.90}
}

@article{Jena2019PerovskiteReview,
  title = {{Halide Perovskite Photovoltaics: Background, Status, and Future Prospects}},
  author = {Jena, Ajay Kumar and Kulkarni, Ashish and Miyasaka, Tsutomu},
  journal = {Chem. Rev.},
  volume = {119},
  number = {5},
  pages = {3036--3103},
  year = {2019},
  doi = {10.1021/acs.chemrev.8b00539}
}

@article{Dresselhaus1955SOC,
  title = {{Spin-Orbit Coupling Effects in Zinc Blende Structures}},
  author = {Dresselhaus, G.},
  journal = {Phys. Rev.},
  volume = {100},
  issue = {2},
  pages = {580--586},
  numpages = {0},
  year = {1955},
  month = {Oct},
  publisher = {American Physical Society},
  doi = {10.1103/PhysRev.100.580},
  url = {https://link.aps.org/doi/10.1103/PhysRev.100.580}
}

@article{Rashba1984,
  title={{Properties of a 2D electron gas with lifted spectral degeneracy}},
  author={Bychkov, {\relax Yu} A and Rashba, {\'E} I},
  journal={JETP Lett.},
  volume={39},
  pages={78},
  year={1984},
  url={http://jetpletters.ru/ps/1264/article_19121.pdf}
}

@article{LuttingerKohn1955,
  title = {{Motion of Electrons and Holes in Perturbed Periodic Fields}},
  author = {Luttinger, J. M. and Kohn, W.},
  journal = {Phys. Rev.},
  volume = {97},
  issue = {4},
  pages = {869--883},
  numpages = {0},
  year = {1955},
  month = {Feb},
  publisher = {American Physical Society},
  doi = {10.1103/PhysRev.97.869},
  url = {https://link.aps.org/doi/10.1103/PhysRev.97.869}
}

@article{pikus1962new,
  title={{A new method of calculating the energy spectrum of carriers in semiconductors. I. Neglecting spin-orbit interaction}},
  author={Pikus, G. E.},
  journal={J. Exp. Theor. Phys.},
  volume={14},
  pages={898},
  year={1962},
  url={http://jetp.ras.ru/cgi-bin/dn/e_014_04_0898.pdf}
}

@article{Xiao2012,
  title = {{Coupled Spin and Valley Physics in Monolayers of ${\mathrm{MoS}}_{2}$ and Other Group-VI Dichalcogenides}},
  author = {Xiao, Di and Liu, Gui-Bin and Feng, Wanxiang and Xu, Xiaodong and Yao, Wang},
  journal = {Phys. Rev. Lett.},
  volume = {108},
  issue = {19},
  pages = {196802},
  numpages = {5},
  year = {2012},
  month = {May},
  publisher = {American Physical Society},
  doi = {10.1103/PhysRevLett.108.196802},
  url = {https://link.aps.org/doi/10.1103/PhysRevLett.108.196802}
}

@article{Splendiani2010,
  title = {{Emerging Photoluminescence in Monolayer MoS$_2$}},
  volume = {10},
  ISSN = {1530-6992},
  url = {http://dx.doi.org/10.1021/nl903868w},
  DOI = {10.1021/nl903868w},
  number = {4},
  journal = {Nano Lett.},
  publisher = {American Chemical Society (ACS)},
  author = {Splendiani,  Andrea and Sun,  Liang and Zhang,  Yuanbo and Li,  Tianshu and Kim,  Jonghwan and Chim,  Chi-Yung and Galli,  Giulia and Wang,  Feng},
  year = {2010},
  month = mar,
  pages = {1271-1275}
}

@Article{Kane1957,
  author    = {Evan O. Kane},
  journal   = {J. Phys. Chem. Solids},
  title     = {{Band structure of indium antimonide}},
  year      = {1957},
  month     = jan,
  number    = {4},
  pages     = {249--261},
  volume    = {1},
  doi       = {10.1016/0022-3697(57)90013-6},
  fjournal  = {Journal of Physics and Chemistry of Solids},
  publisher = {Elsevier {BV}},
  url       = {https://doi.org/10.1016/0022-3697(57)90013-6},
}

@Article{Kane1956,
  author    = {E. O. Kane},
  journal   = {J. Phys. Chem. Solids},
  title     = {{Energy band structure in p-type germanium and silicon}},
  year      = {1956},
  month     = sep,
  number    = {1-2},
  pages     = {82--99},
  volume    = {1},
  doi       = {10.1016/0022-3697(56)90014-2},
  publisher = {Elsevier {BV}},
}

@article{Felipe2020SitePermutation,
  title = {{High-degeneracy points protected by site-permutation symmetries}},
  author = {Crasto de Lima, F. and Ferreira, G. J.},
  journal = {Phys. Rev. B},
  volume = {101},
  issue = {4},
  pages = {041107},
  numpages = {6},
  year = {2020},
  month = {Jan},
  publisher = {American Physical Society},
  doi = {10.1103/PhysRevB.101.041107},
  url = {https://link.aps.org/doi/10.1103/PhysRevB.101.041107}
}

@article{Li2009,
  title = {{Topological Anderson insulator}},
  author = {Li, Jian and Chu, Rui-Lin and Jain, J. K. and Shen, Shun-Qing},
  journal = {Phys. Rev. Lett.},
  volume = {102},
  issue = {13},
  pages = {136806},
  numpages = {4},
  year = {2009},
  month = {Apr},
  publisher = {American Physical Society},
  doi = {10.1103/PhysRevLett.102.136806},
  url = {https://link.aps.org/doi/10.1103/PhysRevLett.102.136806}
}

@article{Groth2009,
  title = {{Theory of the topological Anderson insulator}},
  author = {Groth, C. W. and Wimmer, M. and Akhmerov, A. R. and Tworzyd\l{}o, J. and Beenakker, C. W. J.},
  journal = {Phys. Rev. Lett.},
  volume = {103},
  issue = {19},
  pages = {196805},
  numpages = {4},
  year = {2009},
  month = {Nov},
  publisher = {American Physical Society},
  doi = {10.1103/PhysRevLett.103.196805},
  url = {https://link.aps.org/doi/10.1103/PhysRevLett.103.196805}
}

@article{Sttzer2018,
  title = {{Photonic topological Anderson insulators}},
  volume = {560},
  ISSN = {1476-4687},
  url = {http://dx.doi.org/10.1038/s41586-018-0418-2},
  DOI = {10.1038/s41586-018-0418-2},
  number = {7719},
  journal = {Nature},
  publisher = {Springer Science and Business Media LLC},
  author = {St\"{u}tzer,  Simon and Plotnik,  Yonatan and Lumer,  Yaakov and Titum,  Paraj and Lindner,  Netanel H. and Segev,  Mordechai and Rechtsman,  Mikael C. and Szameit,  Alexander},
  year = {2018},
  month = aug,
  pages = {461}
}

@article{Zhang2022,
  title = {{Topological Anderson insulator via disorder-recovered average symmetry}},
  volume = {106},
  ISSN = {2469-9969},
  pages = {195304},
  url = {http://dx.doi.org/10.1103/PhysRevB.106.195304},
  DOI = {10.1103/physrevb.106.195304},
  number = {19},
  journal = {Phys. Rev. B},
  publisher = {American Physical Society (APS)},
  author = {Zhang,  Jie and Zhang,  Zhi-Qiang and Cheng,  Shu-guang and Jiang,  Hua},
  year = {2022},
  month = nov 
}

@Article{Bryan2024TAI,
  author = {Assun\c{c}ão,  Bryan D. and Ferreira,  Gerson J. and Lewenkopf,  Caio H.},
  title = {{Phase transitions and scale invariance in topological Anderson insulators}},
  doi       = {10.1103/PhysRevB.109.L201102},
  issue     = {20},
  pages     = {L201102},
  url       = {https://link.aps.org/doi/10.1103/PhysRevB.109.L201102},
  volume    = {109},
  journal   = {Phys. Rev. B},
  month     = {May},
  numpages  = {7},
  publisher = {American Physical Society},
  year      = {2024},
}

@article{Caio2019,
  title = {{Topological marker currents in Chern insulators}},
  volume = {15},
  ISSN = {1745-2481},
  url = {http://dx.doi.org/10.1038/s41567-018-0390-7},
  DOI = {10.1038/s41567-018-0390-7},
  number = {3},
  journal = {	Nat. Phys.},
  publisher = {Springer Science and Business Media LLC},
  author = {Caio,  M. D. and M\"{o}ller,  G. and Cooper,  N. R. and Bhaseen,  M. J.},
  year = {2019},
  month = jan,
  pages = {257}
}

@article{Varjas2020,
  title = {{Computation of topological phase diagram of disordered ${\mathrm{Pb}}_{1\ensuremath{-}x}{\mathrm{Sn}}_{x}\mathrm{Te}$ using the kernel polynomial method}},
  author = {Varjas, D\'aniel and Fruchart, Michel and Akhmerov, Anton R. and Perez-Piskunow, Pablo M.},
  journal = {Phys. Rev. Res.},
  volume = {2},
  issue = {1},
  pages = {013229},
  numpages = {11},
  year = {2020},
  month = {Feb},
  publisher = {American Physical Society},
  doi = {10.1103/PhysRevResearch.2.013229},
  url = {https://link.aps.org/doi/10.1103/PhysRevResearch.2.013229}
}

@article{Chiu2016Classification,
  title = {{Classification of topological quantum matter with symmetries}},
  author = {Chiu, Ching-Kai and Teo, Jeffrey C. Y. and Schnyder, Andreas P. and Ryu, Shinsei},
  journal = {Rev. Mod. Phys.},
  volume = {88},
  issue = {3},
  pages = {035005},
  numpages = {63},
  year = {2016},
  month = {Aug},
  publisher = {American Physical Society},
  doi = {10.1103/RevModPhys.88.035005},
  url = {https://link.aps.org/doi/10.1103/RevModPhys.88.035005}
}

@article{AltlandZirnbauer1997TenFoldWay,
  title = {{Nonstandard symmetry classes in mesoscopic normal-superconducting hybrid structures}},
  author = {Altland, Alexander and Zirnbauer, Martin R.},
  journal = {Phys. Rev. B},
  volume = {55},
  issue = {2},
  pages = {1142--1161},
  numpages = {0},
  year = {1997},
  month = {Jan},
  publisher = {American Physical Society},
  doi = {10.1103/PhysRevB.55.1142},
  url = {https://link.aps.org/doi/10.1103/PhysRevB.55.1142}
}

@article{Smejkal2022Altermagnetism,
  title = {{Beyond Conventional Ferromagnetism and Antiferromagnetism: A Phase with Nonrelativistic Spin and Crystal Rotation Symmetry}},
  author = {\ifmmode \check{S}\else \v{S}\fi{}mejkal, Libor and Sinova, Jairo and Jungwirth, Tomas},
  journal = {Phys. Rev. X},
  volume = {12},
  issue = {3},
  pages = {031042},
  numpages = {16},
  year = {2022},
  month = {Sep},
  publisher = {American Physical Society},
  doi = {10.1103/PhysRevX.12.031042},
  url = {https://link.aps.org/doi/10.1103/PhysRevX.12.031042}
}

@article{Smejkal2022EmergingAltermagnetism,
  title = {{Emerging Research Landscape of Altermagnetism}},
  author = {\ifmmode \check{S}\else \v{S}\fi{}mejkal, Libor and Sinova, Jairo and Jungwirth, Tomas},
  journal = {Phys. Rev. X},
  volume = {12},
  issue = {4},
  pages = {040501},
  numpages = {27},
  year = {2022},
  month = {Dec},
  publisher = {American Physical Society},
  doi = {10.1103/PhysRevX.12.040501},
  url = {https://link.aps.org/doi/10.1103/PhysRevX.12.040501}
}

@Article{Krempask2024Altermagnetic,
  author    = {Krempaský, J. and Šmejkal, L. and D'Souza, S. W. and Hajlaoui, M. and Springholz, G. and Uhlířová, K. and Alarab, F. and Constantinou, P. C. and Strocov, V. and Usanov, D. and Pudelko, W. R. and González-Hernández, R. and Birk Hellenes, A. and Jansa, Z. and Reichlová, H. and Šobáň, Z. and Gonzalez Betancourt, R. D. and Wadley, P. and Sinova, J. and Kriegner, D. and Minár, J. and Dil, J. H. and Jungwirth, T.},
  title     = {{Altermagnetic lifting of Kramers spin degeneracy}},
  doi       = {10.1038/s41586-023-06907-7},
  issn      = {1476-4687},
  number    = {7999},
  pages     = {517},
  url       = {http://dx.doi.org/10.1038/s41586-023-06907-7},
  volume    = {626},
  journal   = {Nature},
  month     = feb,
  publisher = {Springer Science and Business Media LLC},
  year      = {2024},
}

@article{Osumi2024Altermagnetic,
  title = {{Observation of a giant band splitting in altermagnetic MnTe}},
  author = {Osumi, T. and Souma, S. and Aoyama, T. and Yamauchi, K. and Honma, A. and Nakayama, K. and Takahashi, T. and Ohgushi, K. and Sato, T.},
  journal = {Phys. Rev. B},
  volume = {109},
  issue = {11},
  pages = {115102},
  numpages = {8},
  year = {2024},
  month = {Mar},
  publisher = {American Physical Society},
  doi = {10.1103/PhysRevB.109.115102},
  url = {https://link.aps.org/doi/10.1103/PhysRevB.109.115102}
}

@book{koster1963properties,
  title={{The Properties of the Thirty-Two Point Groups}},
  author={Koster, George F and Dimmock, John O and Wheeler, Robert G and Statz, Hermann},
  year={1963},
  publisher={MIT Press},
  address={Cambridge, Mass.},
  isbn={9780262110105},
  url = {https://mitpress.mit.edu/9780262110105/the-properties-of-the-thirty-two-point-groups/}
}

@Article{Aroyo2011Bilbao1,
  author            = {Aroyo, M. I. and Perez-Mato, J. M. and Orobengoa, D. and Tasci, E. and De La Flor, G. and Kirov, A.},
  journal           = {Bulg. Chem. Commun.},
  title = {{Crystallography online: Bilbao crystallographic server}},
  year              = {2011},
  number            = {2},
  pages             = {183},
  volume            = {43},
  fjournal          = {Bulgarian Chemical Communications},
  publication_stage = {Final},
  source            = {Scopus},
  type              = {Article},
  url               = {https://www.scopus.com/pages/publications/80955140447},
}

@Article{Aroyo2006Bilbao2,
  author    = {Aroyo, Mois Ilia and Perez-Mato, Juan Manuel and Capillas, Cesar and Kroumova, Eli and Ivantchev, Svetoslav and Madariaga, Gotzon and Kirov, Asen and Wondratschek, Hans},
  title = {{Bilbao Crystallographic Server: I. Databases and crystallographic computing programs}},
  doi       = {10.1524/zkri.2006.221.1.15},
  issn      = {2194-4946},
  number    = {1},
  pages     = {15},
  url       = {http://dx.doi.org/10.1524/zkri.2006.221.1.15},
  volume    = {221},
  journal   = {Z. Kristallogr. Cryst. Mater.},
  month     = jan,
  publisher = {Walter de Gruyter GmbH},
  year      = {2006},
}

@Article{Aroyo2006Bilbao3,
  author    = {Aroyo, Mois I. and Kirov, Asen and Capillas, Cesar and Perez-Mato, J. M. and Wondratschek, Hans},
  title = {{Bilbao Crystallographic Server. II. Representations of crystallographic point groups and space groups}},
  doi       = {10.1107/s0108767305040286},
  issn      = {0108-7673},
  number    = {2},
  pages     = {115},
  url       = {http://dx.doi.org/10.1107/S0108767305040286},
  volume    = {62},
  fjournal  = {Acta Crystallographica Section A Foundations of Crystallography},
  journal   = {Acta Crystallogr. A},
  month     = mar,
  publisher = {International Union of Crystallography (IUCr)},
  year      = {2006},
}

@article{Fu2020Wurtzite,
  title = {{Spin-orbit coupling in wurtzite heterostructures}},
  author = {Fu, Jiyong and Penteado, Poliana H. and Candido, Denis R. and Ferreira, G. J. and Pires, D. P. and Bernardes, E. and Egues, J. C.},
  journal = {Phys. Rev. B},
  volume = {101},
  issue = {13},
  pages = {134416},
  numpages = {27},
  year = {2020},
  month = {Apr},
  publisher = {American Physical Society},
  doi = {10.1103/PhysRevB.101.134416},
  url = {https://link.aps.org/doi/10.1103/PhysRevB.101.134416}
}

@Article{Hopfdeld1960,
  author    = {Hopfdeld, J.J.},
  title     = {{Fine structure in the optical absorption edge of anisotropic crystals}},
  doi       = {10.1016/0022-3697(60)90105-0},
  issn      = {0022-3697},
  number    = {1-2},
  pages     = {97-107},
  url       = {http://dx.doi.org/10.1016/0022-3697(60)90105-0},
  volume    = {15},
  fjournal  = {Journal of Physics and Chemistry of Solids},
  journal   = {J. Phys. Chem. Solids},
  month     = aug,
  publisher = {Elsevier BV},
  year      = {1960},
}

@article{Voon1996,
  title = {{Terms linear in k in the band structure of wurtzite-type semiconductors}},
  author = {Lew Yan Voon, L. C. and Willatzen, M. and Cardona, M. and Christensen, N. E.},
  journal = {Phys. Rev. B},
  volume = {53},
  issue = {16},
  pages = {10703--10714},
  numpages = {0},
  year = {1996},
  month = {Apr},
  publisher = {American Physical Society},
  doi = {10.1103/PhysRevB.53.10703},
  url = {https://link.aps.org/doi/10.1103/PhysRevB.53.10703}
}

@Article{ChuangChang1996,
  author    = {Chuang, S. L. and Chang, C. S.},
  title     = {{k\ifmmode\cdot\else\textperiodcentered\fi{}p method for strained wurtzite semiconductors}},
  doi       = {10.1103/PhysRevB.54.2491},
  issue     = {4},
  pages     = {2491--2504},
  url       = {https://link.aps.org/doi/10.1103/PhysRevB.54.2491},
  volume    = {54},
  journal   = {Phys. Rev. B},
  month     = {Jul},
  numpages  = {0},
  publisher = {American Physical Society},
  year      = {1996},
}

@Article{Gutsche1967,
  author    = {Gutsche, E. and Jahne, E.},
  title     = {{Spin‐Orbit Splitting of the Valence Band of Wurtzite Type Crystals}},
  doi       = {10.1002/pssb.19670190235},
  issn      = {1521-3951},
  number    = {2},
  pages     = {823-832},
  url       = {http://dx.doi.org/10.1002/pssb.19670190235},
  volume    = {19},
  fjournal  = {physica status solidi (b)},
  journal   = {Phys. Status Solidi B},
  month     = jan,
  publisher = {Wiley},
  year      = {1967},
}

@article{Dugdale2000wurtzite,
  title = {{Direct calculation of $\mathbf{k}\ensuremath{\cdot}\mathbf{p}$ parameters for wurtzite AlN, GaN, and InN}},
  author = {Dugdale, D. J. and Brand, S. and Abram, R. A.},
  journal = {Phys. Rev. B},
  volume = {61},
  issue = {19},
  pages = {12933--12938},
  numpages = {0},
  year = {2000},
  month = {May},
  publisher = {American Physical Society},
  doi = {10.1103/PhysRevB.61.12933},
  url = {https://link.aps.org/doi/10.1103/PhysRevB.61.12933}
}

@Article{faria2016realistic,
  author    = {Faria Junior, Paulo E and Campos, Tiago and Bastos, Carlos MO and Gmitra, Martin and Fabian, Jaroslav and Sipahi, Guilherme M},
  title     = {{Realistic multiband k\ifmmode\cdot\else\textperiodcentered\fi{}p approach from ab initio and spin-orbit coupling effects of InAs and InP in wurtzite phase}},
  number    = {23},
  pages     = {235204},
  volume    = {93},
  journal   = {Phys. Rev. B},
  publisher = {APS},
  year      = {2016},
  doi = {10.1103/PhysRevB.93.235204}
}

@article{Langer1970,
  title = {{Spin Exchange in Excitons,  the Quasicubic Model and Deformation Potentials in II-VI Compounds}},
  volume = {2},
  ISSN = {0556-2805},
  url = {http://dx.doi.org/10.1103/PhysRevB.2.4005},
  DOI = {10.1103/physrevb.2.4005},
  number = {10},
  journal = {Phys. Rev. B},
  publisher = {American Physical Society (APS)},
  author = {Langer,  D. W. and Euwema,  R. N. and Era,  Koh and Koda,  Takao},
  year = {1970},
  month = nov,
  pages = {4005}
}

@Article{Suzuki1995,
  author    = {Suzuki, Masakatsu and Uenoyama, Takeshi and Yanase, Akira},
  title     = {{First-principles calculations of effective-mass parameters of AlN and GaN}},
  doi       = {10.1103/PhysRevB.52.8132},
  issue     = {11},
  pages     = {8132--8139},
  url       = {https://link.aps.org/doi/10.1103/PhysRevB.52.8132},
  volume    = {52},
  journal   = {Phys. Rev. B},
  month     = {Sep},
  numpages  = {0},
  publisher = {American Physical Society},
  year      = {1995},
}

@article{Yamaguchi1998,
  title = {{Determination of valence band splitting parameters in GaN}},
  volume = {83},
  ISSN = {1089-7550},
  url = {http://dx.doi.org/10.1063/1.367217},
  DOI = {10.1063/1.367217},
  number = {8},
  journal = {	J. Appl. Phys.},
  publisher = {AIP Publishing},
  author = {Yamaguchi,  A. A. and Mochizuki,  Y. and Sunakawa,  H. and Usui,  A.},
  year = {1998},
  month = apr,
  pages = {4542}
}

@article{Lambrecht2002,
  title = {{Valence-band ordering and magneto-optic exciton fine structure in ZnO}},
  volume = {65},
  ISSN = {1095-3795},
  url = {http://dx.doi.org/10.1103/PhysRevB.65.075207},
  DOI = {10.1103/physrevb.65.075207},
  number = {7},
  journal = {Phys. Rev. B},
  publisher = {American Physical Society (APS)},
  author = {Lambrecht,  Walter R. L. and Rodina,  Anna V. and Limpijumnong,  Sukit and Segall,  B. and Meyer,  Bruno K.},
  year = {2002},
  month = jan,
  pages = 075207
}

@Article{Guo2022,
  author  = {Guo, Chunyu and Hu, Lunhui and Putzke, Carsten and Diaz, Jonas and Huang, Xiangwei and Manna, Kaustuv and Fan, Feng-Ren and Shekhar, Chandra and Sun, Yan and Felser, Claudia and Liu, Chaoxing and Bernevig, B. Andrei and Moll, Philip J. W.},
  title   = {{Quasi-symmetry-protected topology in a semi-metal}},
  doi     = {10.1038/s41567-022-01604-0},
  number  = {7},
  pages   = {813-818},
  volume  = {18},
  journal = {Nat. Phys.},
  year    = {2022},
}

@article{Tao2023,
  title = {{Rashba-like spin splitting around non-time-reversal-invariant momenta}},
  volume = {107},
  DOI = {10.1103/physrevb.107.235138},
  number = {23},
  journal = {Phys. Rev. B},
  author = {Tao,  L. L. and Li,  Jiayu and Liu,  Yuntian and Wang,  Xianjie and Sui,  Yu and Song,  Bo and Zhuravlev,  M. Ye. and Liu,  Qihang},
  year = {2023},
  pages = {235138}
}

@article{Hu2023Hierarchy,
  title = {{Hierarchy of quasisymmetries and degeneracies in the CoSi family of chiral crystal materials}},
  volume = {107},
  DOI = {10.1103/PhysRevB.107.125145},
  number = {31},
  journal = {Phys. Rev. B},  
  author = {Hu, Lun-Hui and Guo, Chunyu and Sun, Yan and Felser, Claudia and Elcoro, Luis and Moll, Philip J. W. and Liu, Chao-Xing and Bernevig, B. Andrei},
  year = {2023},
  pages = {125145}
}

@article{PRL2024,
  title = {{Group Theory on Quasisymmetry and Protected Near Degeneracy}},
  volume = {133},
  DOI = {10.1103/PhysRevLett.133.026402},  
  number = {7},
  journal = {Phys. Rev. Lett.},
  author = {Li, Jiayu and Zhang, Ao and Liu, Yuntian and Liu, Qihang},
  year = {2024},
  pages = {026402}
}

@article{Zhang2025,
  title = {{Topological charge quadrupole protected by spin-orbit U(1) quasi-symmetry in antiferromagnet NdBiPt}},
  author={Zhang, Ao and Chen, Xiaobing and Li, Jiayu and Liu, Pengfei and Liu, Yuntian and Liu, Qihang},
  journal={Newton},
  volume={1},
  number={2},
  year={2025},
  publisher={Elsevier},
  doi = {/10.1016/j.newton.2025.100010},
  pages = {100010}
}

@Article{Liu2024,
  author    = {Liu, Lu and Liu, Yuntian and Li, Jiayu and Wu, Hua and Liu, Qihang},
  title = {{Quantum spin Hall effect protected by spin U(1) quasisymmetry}},
  doi       = {10.1103/physrevb.110.l161104},
  issue     = {16},
  number    = {16},
  pages     = {L161104},
  url       = {https://link.aps.org/doi/10.1103/PhysRevB.110.L161104},
  volume    = {110},
  journal   = {Phys. Rev. B},
  month     = {Oct},
  numpages  = {7},
  publisher = {American Physical Society},
  year      = {2024},
}

@article{Liu2024_orbital,
  title = {{Orbital doublet driven even-spin Chern insulators}},
  author = {Liu, Lu and Liu, Yuntian and Li, Jiayu and Wu, Hua and Liu, Qihang},
  journal = {Phys. Rev. B},
  volume = {110},
  issue = {3},
  pages = {035161},
  numpages = {8},
  year = {2024},
  month = {Jul},
  publisher = {American Physical Society},
  doi = {10.1103/PhysRevB.110.035161},
  url = {https://link.aps.org/doi/10.1103/PhysRevB.110.035161}
}

@Unpublished{li2026quasisymmetry,
  author        = {Jiayu Li and Feng-Ren Fan and Wang Yao},
  title = {{Quasisymmetry Enriched Gapless Criticality at Chern Insulator Transitions}},
  eprint        = {2601.15011},
  url           = {https://arxiv.org/abs/2601.15011},
  archiveprefix = {arXiv},
  primaryclass  = {cond-mat.mtrl-sci},
  year          = {2026},
}

@article{Tao2023_ising,
  title = {{Quasisymmetry protected Ising spin-orbit coupling}},
  volume = {107},
  ISSN = {2469-9969},
  url = {http://dx.doi.org/10.1103/PhysRevB.107.155412},
  DOI = {10.1103/physrevb.107.155412},
  pages = {155412},
  number = {15},
  journal = {Phys. Rev. B},
  publisher = {American Physical Society (APS)},
  author = {Tao,  L. L. and Zhang,  Qin and Li,  Huinan and Wang,  Xianjie and Wang,  Yi and Sui,  Yu and Song,  Bo and Zhuravlev,  M. Ye.},
  year = {2023},
  month = apr 
}

@article{Roig2025Altermagnet,
  title = {{Quasisymmetry-Constrained Spin Ferromagnetism in Altermagnets}},
  author = {Roig, Merc\`e and Yu, Yue and Ekman, Rune C. and Kreisel, Andreas and Andersen, Brian M. and Agterberg, Daniel F.},
  journal = {Phys. Rev. Lett.},
  volume = {135},
  issue = {1},
  pages = {016703},
  numpages = {8},
  year = {2025},
  month = {Jul},
  publisher = {American Physical Society},
  doi = {10.1103/839n-rckn},
  url = {https://link.aps.org/doi/10.1103/839n-rckn}
}

@article{Kang2024Majorna,
  title = {{Subsymmetry protected topology in topological insulators and superconductors}},
  author = {Kang, Myungjun and Lee, Mingyu and Cheon, Sangmo},
  journal = {Phys. Rev. Res.},
  volume = {6},
  issue = {3},
  pages = {033323},
  numpages = {9},
  year = {2024},
  month = {Sep},
  publisher = {American Physical Society},
  doi = {10.1103/PhysRevResearch.6.033323},
  url = {https://link.aps.org/doi/10.1103/PhysRevResearch.6.033323}
}

@Article{Ren2021,
  author    = {Ren, Jie and Liang, Chenguang and Fang, Chen},
  title = {{Quasisymmetry Groups and Many-Body Scar Dynamics}},
  doi       = {10.1103/physrevlett.126.120604},
  issn      = {1079-7114},
  number    = {12},
  url       = {http://dx.doi.org/10.1103/PhysRevLett.126.120604},
  pages = {120604},
  volume    = {126},
  fjournal  = {Physical Review Letters},
  journal   = {Phys. Rev. Lett.},
  month     = mar,
  publisher = {American Physical Society (APS)},
  year      = {2021},
}

@article{Ren2024,
  title = {{Quasi-Nambu-Goldstone modes in many-body scar models}},
  volume = {110},
  ISSN = {2469-9969},
  url = {http://dx.doi.org/10.1103/PhysRevB.110.245101},
  pages = {245101},
  DOI = {10.1103/physrevb.110.245101},
  number = {24},
  journal = {Phys. Rev. B},
  publisher = {American Physical Society (APS)},
  author = {Ren,  Jie and Wang,  Yu-Peng and Fang,  Chen},
  year = {2024},
  month = dec 
}

@article{Yaji2019,
  title = {{Coexistence of Two Types of Spin Splitting Originating from Different Symmetries}},
  volume = {122},
  DOI = {10.1103/physrevlett.122.126403},
  number = {12},
  journal = {Phys. Rev. Lett.},
  author = {Yaji,  Koichiro and Visikovskiy,  Anton and Iimori,  Takushi and Kuroda,  Kenta and Hayashi,  Singo and Kajiwara,  Takashi and Tanaka,  Satoru and Komori,  Fumio and Shin,  Shik},
  year = {2019},
  pages = {126403}
}

@article{Woniak2020,
  title = {{Exciton $g$ factors of van der Waals heterostructures from first-principles calculations}},
  author = {Wo\ifmmode \acute{z}\else \'{z}\fi{}niak, Tomasz and Faria Junior, Paulo E. and Seifert, Gotthard and Chaves, Andrey and Kunstmann, Jens},
  journal = {Phys. Rev. B},
  volume = {101},
  issue = {23},
  pages = {235408},
  numpages = {11},
  year = {2020},
  month = {Jun},
  publisher = {American Physical Society},
  doi = {10.1103/PhysRevB.101.235408},
  url = {https://link.aps.org/doi/10.1103/PhysRevB.101.235408}
}

@article{Glazov2024,
  title = {{Excitons in two-dimensional materials and heterostructures: Optical and magneto-optical properties}},
  volume = {49},
  ISSN = {1938-1425},
  url = {http://dx.doi.org/10.1557/s43577-024-00754-1},
  DOI = {10.1557/s43577-024-00754-1},
  number = {9},
  journal = {MRS Bull.},
  publisher = {Springer Science and Business Media LLC},
  author = {Glazov,  Mikhail and Arora,  Ashish and Chaves,  Andrey and Gobato,  Yara Galvão},
  year = {2024},
  month = aug,
  pages = {899}
}

@article{Deb2024,
  title = {{Excitonic signatures of ferroelectric order in parallel-stacked MoS$_2$}},
  volume = {15},
  pages = {7595},
  ISSN = {2041-1723},
  url = {http://dx.doi.org/10.1038/s41467-024-52011-3},
  DOI = {10.1038/s41467-024-52011-3},
  number = {1},
  journal = {	Nat. Commun.},
  publisher = {Springer Science and Business Media LLC},
  author = {Deb,  Swarup and Krause,  Johannes and Faria Junior,  Paulo E. and Kempf,  Michael Andreas and Schwartz,  Rico and Watanabe,  Kenji and Taniguchi,  Takashi and Fabian,  Jaroslav and Korn,  Tobias},
  year = {2024},
  month = aug 
}

@article{FariaJr2025InterGFactor,
  title = {{Generalized many-body exciton $g$ factors: Magnetic hybridization and nonmonotonic Rydberg series in monolayer ${\mathrm{WSe}}_{2}$}},
  author = {Faria Junior, Paulo E. and Hernang\'omez-P\'erez, Daniel and Amit, Tomer and Fabian, Jaroslav and Refaely-Abramson, Sivan},
  journal = {Phys. Rev. B},
  volume = {112},
  issue = {24},
  pages = {L241404},
  numpages = {9},
  year = {2025},
  month = {Dec},
  publisher = {American Physical Society},
  doi = {10.1103/c5y7-w9tk},
  url = {https://link.aps.org/doi/10.1103/c5y7-w9tk}
}

@article{Kresse1993,
  title = {{Ab initio molecular dynamics for liquid metals}},
  author = {Kresse, G. and Hafner, J.},
  journal = {Phys. Rev. B},
  volume = {47},
  issue = {1},
  pages = {558--561},
  numpages = {0},
  year = {1993},
  month = {Jan},
  publisher = {American Physical Society},
  doi = {10.1103/PhysRevB.47.558},
  url = {https://link.aps.org/doi/10.1103/PhysRevB.47.558}
}

@article{Kresse1996,
  title = {{Efficient iterative schemes for ab initio total-energy calculations using a plane-wave basis set}},
  author = {Kresse, G. and Furthm\"uller, J.},
  journal = {Phys. Rev. B},
  volume = {54},
  issue = {16},
  pages = {11169--11186},
  numpages = {0},
  year = {1996},
  month = {Oct},
  publisher = {American Physical Society},
  doi = {10.1103/PhysRevB.54.11169},
  url = {https://link.aps.org/doi/10.1103/PhysRevB.54.11169}
}

@article{Perdew1996,
  title = {{Generalized Gradient Approximation Made Simple}},
  author = {Perdew, John P. and Burke, Kieron and Ernzerhof, Matthias},
  journal = {Phys. Rev. Lett.},
  volume = {77},
  issue = {18},
  pages = {3865--3868},
  numpages = {0},
  year = {1996},
  month = {Oct},
  publisher = {American Physical Society},
  doi = {10.1103/PhysRevLett.77.3865},
  url = {https://link.aps.org/doi/10.1103/PhysRevLett.77.3865}
}

@Article{Giannozzi2009,
  author    = {Paolo Giannozzi and Stefano Baroni and Nicola Bonini and Matteo Calandra and Roberto Car and Carlo Cavazzoni and Davide Ceresoli and Guido L Chiarotti and Matteo Cococcioni and Ismaila Dabo and Andrea Dal Corso and Stefano de Gironcoli and Stefano Fabris and Guido Fratesi and Ralph Gebauer and Uwe Gerstmann and Christos Gougoussis and Anton Kokalj and Michele Lazzeri and Layla Martin-Samos and Nicola Marzari and Francesco Mauri and Riccardo Mazzarello and Stefano Paolini and Alfredo Pasquarello and Lorenzo Paulatto and Carlo Sbraccia and Sandro Scandolo and Gabriele Sclauzero and Ari P Seitsonen and Alexander Smogunov and Paolo Umari and Renata M Wentzcovitch},
  journal   = {J. Phys.: Condens. Matter},
  title     = {{QUANTUM} {ESPRESSO}: a modular and open-source software project for quantum simulations of materials},
  year      = {2009},
  month     = sep,
  number    = {39},
  pages     = {395502},
  volume    = {21},
  doi       = {10.1088/0953-8984/21/39/395502},
  publisher = {{IOP} Publishing},
}

@Article{Giannozzi2017,
  author    = {P Giannozzi and O Andreussi and T Brumme and O Bunau and M Buongiorno Nardelli and M Calandra and R Car and C Cavazzoni and D Ceresoli and M Cococcioni and N Colonna and I Carnimeo and A Dal Corso and S de Gironcoli and P Delugas and R A DiStasio and A Ferretti and A Floris and G Fratesi and G Fugallo and R Gebauer and U Gerstmann and F Giustino and T Gorni and J Jia and M Kawamura and H-Y Ko and A Kokalj and E K{\"{u}}{\c{c}}{\"{u}}kbenli and M Lazzeri and M Marsili and N Marzari and F Mauri and N L Nguyen and H-V Nguyen and A Otero-de-la-Roza and L Paulatto and S Ponc{\'{e}} and D Rocca and R Sabatini and B Santra and M Schlipf and A P Seitsonen and A Smogunov and I Timrov and T Thonhauser and P Umari and N Vast and X Wu and S Baroni},
  journal   = {J. Phys.: Condens. Matter},
  title     = {Advanced capabilities for materials modelling with Quantum {ESPRESSO}},
  year      = {2017},
  month     = oct,
  number    = {46},
  pages     = {465901},
  volume    = {29},
  doi       = {10.1088/1361-648x/aa8f79},
  publisher = {{IOP} Publishing},
}

@article{Cassiano2024,
   title = {{DFT2kp: Effective kp models from ab-initio data}},
   ISSN={2949-804X},
   url={http://dx.doi.org/10.21468/SciPostPhysCodeb.25},
   DOI={10.21468/scipostphyscodeb.25},
   journal={SciPost Physics Codebases},
   publisher={Stichting SciPost},
   author={V. Cassiano, João Victor and de Lelis Araújo, Augusto and Faria Junior, Paulo E. and J. Ferreira, Gerson},
   year={2024},
   pages = {25},
   volume = {},
   month=feb 
}

@Article{IrRep2022,
  author    = {Iraola, Mikel and Mañes, Juan L. and Bradlyn, Barry and Horton, Matthew K. and Neupert, Titus and Vergniory, Maia G. and Tsirkin, Stepan S.},
  title     = {{IrRep: Symmetry eigenvalues and irreducible representations of ab initio band structures}},
  doi       = {10.1016/j.cpc.2021.108226},
  issn      = {0010-4655},
  pages     = {108226},
  url       = {http://dx.doi.org/10.1016/j.cpc.2021.108226},
  volume    = {272},
  fjournal  = {Computer Physics Communications},
  journal   = {Comput. Phys. Commun.},
  month     = mar,
  publisher = {Elsevier BV},
  year      = {2022},
}

@Article{WIEN2k2020,
  author    = {Blaha, Peter and Schwarz, Karlheinz and Tran, Fabien and Laskowski, Robert and Madsen, Georg K. H. and Marks, Laurence D.},
  title = {{WIEN2k: An APW+lo program for calculating the properties of solids}},
  doi       = {10.1063/1.5143061},
  issn      = {1089-7690},
  number    = {7},
  url       = {http://dx.doi.org/10.1063/1.5143061},
  volume    = {152},
  fjournal  = {The Journal of Chemical Physics},
  journal   = {J. Chem. Phys.},
  month     = feb,
  publisher = {AIP Publishing},
  year      = {2020},
  pages = 074101
}

@dataset{PythTB,
    author = {Cole, Trey and Coh, Sinisa and Vanderbilt, David},
    doi = {10.5281/zenodo.12721315},
    license = {GPL-3.0-or-later},
    title = {{Python Tight Binding (PythTB) (Version 2.0.0) [Computer Software]}},
    url = {https://zenodo.org/records/12721315},
    note = {{Zenodo}},
    year = {2025}
}

@Article{Kroumova2001,
  author    = {Kroumova, E. and Aroyo, M. I. and Perez-Mato, J. M. and Ivantchev, S. and Igartua, J. M. and Wondratschek, H.},
  title = {{PSEUDO: a program for a pseudosymmetry search}},
  doi       = {10.1107/s0021889801011852},
  issn      = {0021-8898},
  number    = {6},
  pages     = {783-784},
  url       = {http://dx.doi.org/10.1107/S0021889801011852},
  volume    = {34},
  fjournal  = {Journal of Applied Crystallography},
  journal   = {J. Appl. Crystallogr.},
  month     = nov,
  publisher = {International Union of Crystallography (IUCr)},
  year      = {2001},
}

@Article{Zwart2007,
  author    = {Zwart, Peter H. and Grosse-Kunstleve, Ralf W. and Lebedev, Andrey A. and Murshudov, Garib N. and Adams, Paul D.},
  title = {{Surprises and pitfalls arising from (pseudo)symmetry}},
  doi       = {10.1107/s090744490705531x},
  issn      = {0907-4449},
  number    = {1},
  pages     = {99},
  url       = {http://dx.doi.org/10.1107/S090744490705531X},
  volume    = {64},
  journal   = {Acta Crystallogr. D},
  month     = dec,
  publisher = {International Union of Crystallography (IUCr)},
  year      = {2007},
}

@Article{capillas2011new,
  author      = {Cesar Capillas and Emre Sururi Tasci and Gemma de la Flor and Danel Orobengoa and Juan Manuel Perez-Mato and Mois Ilia Aroyo},
  title = {{A new computer tool at the Bilbao Crystallographic Server to detect and characterize pseudosymmetry}},
  doi         = {doi:10.1524/zkri.2011.1321},
  number      = {2},
  pages       = {186--196},
  url         = {https://doi.org/10.1524/zkri.2011.1321},
  volume      = {226},
  journal     = {Z. Kristallogr. Cryst. Mater.},
  lastchecked = {2026-01-06},
  year        = {2011},
}

@Article{Nolze2023pseudo,
  author    = {Nolze, Gert and Tokarski, Tomasz and Rychlowski, Lukasz},
  title = {{Use of electron backscatter diffraction patterns to determine the crystal lattice. Part 3. Pseudosymmetry}},
  doi       = {10.1107/s1600576723000845},
  issn      = {1600-5767},
  number    = {2},
  pages     = {367},
  url       = {http://dx.doi.org/10.1107/S1600576723000845},
  volume    = {56},
  fjournal  = {Journal of Applied Crystallography},
  journal   = {J. Appl. Crystallogr.},
  month     = feb,
  publisher = {International Union of Crystallography (IUCr)},
  year      = {2023},
}

@Article{mostofi2014updated,
  author    = {Mostofi, Arash A and Yates, Jonathan R and Pizzi, Giovanni and Lee, Young-Su and Souza, Ivo and Vanderbilt, David and Marzari, Nicola},
  title = {{An updated version of wannier90: A tool for obtaining maximally-localised Wannier functions}},
  number    = {8},
  pages     = {2309--2310},
  volume    = {185},
  fjournal  = {Computer Physics Communications},
  journal   = {Comput. Phys. Commun.},
  publisher = {Elsevier},
  year      = {2014},
  doi = {10.1016/j.cpc.2014.05.003},
}

@Article{pizzi2020wannier90,
  author    = {Pizzi, Giovanni and Vitale, Valerio and Arita, Ryotaro and Bl{\"u}gel, Stefan and Freimuth, Frank and G{\'e}ranton, Guillaume and Gibertini, Marco and Gresch, Dominik and Johnson, Charles and Koretsune, Takashi and others},
  title = {{Wannier90 as a community code: new features and applications}},
  number    = {16},
  pages     = {165902},
  volume    = {32},
  fjournal  = {Journal of Physics: Condensed Matter},
  journal   = {J. Phys. Condens. Matter},
  publisher = {IOP Publishing},
  year      = {2020},
  doi = {10.1088/1361-648X/ab51ff}
}

@Article{mostofi2008wannier90,
  author    = {Mostofi, Arash A and Yates, Jonathan R and Lee, Young-Su and Souza, Ivo and Vanderbilt, David and Marzari, Nicola},
  title = {{wannier90: A tool for obtaining maximally-localised Wannier functions}},
  number    = {9},
  pages     = {685},
  volume    = {178},
  journal   = {Comput. Phys. Commun.},
  publisher = {Elsevier},
  year      = {2008},
  doi = {10.1016/j.cpc.2007.11.016}
}

@Article{SLWF2014,
  author    = {Wang, Runzhi and Lazar, Emanuel A. and Park, Hyowon and Millis, Andrew J. and Marianetti, Chris A.},
  title = {{Selectively localized Wannier functions}},
  doi       = {10.1103/physrevb.90.165125},
  issn      = {1550-235X},
  number    = {16},
  url       = {http://dx.doi.org/10.1103/PhysRevB.90.165125},
  volume    = {90},
  fjournal  = {Physical Review B},
  journal   = {Phys. Rev. B},
  month     = oct,
  publisher = {American Physical Society (APS)},
  year      = {2014},
  pages = {165125}
}

@Article{WannierBerri,
  author    = {Tsirkin, Stepan S.},
  title = {{High performance Wannier interpolation of Berry curvature and related quantities with WannierBerri code}},
  doi       = {10.1038/s41524-021-00498-5},
  issn      = {2057-3960},
  number    = {1},
  url       = {http://dx.doi.org/10.1038/s41524-021-00498-5},
  volume    = {7},
  fjournal  = {npj Computational Materials},
  journal   = {NPJ Comput. Mater.},
  month     = feb,
  publisher = {Springer Science and Business Media LLC},
  year      = {2021},
  pages = {33}
}

@article{SAWF2013,
  title = {{Symmetry-adapted Wannier functions in the maximal localization procedure}},
  author = {Sakuma, R.},
  journal = {Phys. Rev. B},
  volume = {87},
  issue = {23},
  pages = {235109},
  numpages = {8},
  year = {2013},
  month = {Jun},
  publisher = {American Physical Society},
  doi = {10.1103/PhysRevB.87.235109},
  url = {https://link.aps.org/doi/10.1103/PhysRevB.87.235109}
}

@article{Schnyder2008TopoSupercond,
  title = {{Classification of topological insulators and superconductors in three spatial dimensions}},
  author = {Schnyder, Andreas P. and Ryu, Shinsei and Furusaki, Akira and Ludwig, Andreas W. W.},
  journal = {Phys. Rev. B},
  volume = {78},
  issue = {19},
  pages = {195125},
  numpages = {22},
  year = {2008},
  month = {Nov},
  publisher = {American Physical Society},
  doi = {10.1103/PhysRevB.78.195125},
  url = {https://link.aps.org/doi/10.1103/PhysRevB.78.195125}
}

@InProceedings{Kitaev2009,
  title = {{Periodic table for topological insulators and superconductors}},
  url = {http://dx.doi.org/10.1063/1.3149495},
  DOI = {10.1063/1.3149495},
  booktitle = {AIP Conference Proceedings},
  publisher = {AIP},
  author = {Kitaev,  Alexei and Lebedev,  Vladimir and Feigel’man,  Mikhail},
  year = {2009},
  pages = {22}
}

@Article{Sato2017,
  author    = {Sato, Masatoshi and Ando, Yoichi},
  title     = {{Topological superconductors: A review}},
  doi       = {10.1088/1361-6633/aa6ac7},
  issn      = {1361-6633},
  number    = {7},
  pages     = {076501},
  url       = {http://dx.doi.org/10.1088/1361-6633/aa6ac7},
  volume    = {80},
  fjournal  = {Reports on Progress in Physics},
  journal   = {Rep. Prog. Phys.},
  month     = may,
  publisher = {IOP Publishing},
  year      = {2017},
}

@Article{Neupert2021,
  author    = {Neupert, Titus and Denner, M. Michael and Yin, Jia-Xin and Thomale, Ronny and Hasan, M. Zahid},
  title     = {{Charge order and superconductivity in Kagome materials}},
  doi       = {10.1038/s41567-021-01404-y},
  issn      = {1745-2481},
  number    = {2},
  pages     = {137},
  url       = {http://dx.doi.org/10.1038/s41567-021-01404-y},
  volume    = {18},
  fjournal  = {Nature Physics},
  journal   = {Nat. Phys.},
  month     = dec,
  publisher = {Springer Science and Business Media LLC},
  year      = {2021},
}

@Article{Zhang2024,
  title = {{Topological superconductivity from unconventional band degeneracy with conventional pairing}},
  volume = {15},
  ISSN = {2041-1723},
  url = {http://dx.doi.org/10.1038/s41467-024-52156-1},
  DOI = {10.1038/s41467-024-52156-1},
  number = {1},
  journal = {Nat. Comm.},
  publisher = {Springer Science and Business Media LLC},
  author = {Zhang,  Zhongyi and Wu,  Zhenfei and Fang,  Chen and Zhang,  Fu-chun and Hu,  Jiangping and Wang,  Yuxuan and Qin,  Shengshan},
  year = {2024},
  month = Sept,
  pages = {7971}
}

@article{KaneMele2005SHEGraphene,
  title = {{Quantum Spin Hall Effect in Graphene}},
  author = {Kane, C. L. and Mele, E. J.},
  journal = {Phys. Rev. Lett.},
  volume = {95},
  issue = {22},
  pages = {226801},
  numpages = {4},
  year = {2005},
  month = {Nov},
  publisher = {American Physical Society},
  doi = {10.1103/PhysRevLett.95.226801},
  url = {https://link.aps.org/doi/10.1103/PhysRevLett.95.226801}
}

@article{KaneMele2005Z2QSHE,
  title = {{${Z}_{2}$ Topological Order and the Quantum Spin Hall Effect}},
  author = {Kane, C. L. and Mele, E. J.},
  journal = {Phys. Rev. Lett.},
  volume = {95},
  issue = {14},
  pages = {146802},
  numpages = {4},
  year = {2005},
  month = {Sep},
  publisher = {American Physical Society},
  doi = {10.1103/PhysRevLett.95.146802},
  url = {https://link.aps.org/doi/10.1103/PhysRevLett.95.146802}
}

@Article{Bernevig2006,
  author    = {Bernevig, B. Andrei and Hughes, Taylor L. and Zhang, Shou-Cheng},
  title = {{Quantum Spin Hall Effect and Topological Phase Transition in HgTe Quantum Wells}},
  doi       = {10.1126/science.1133734},
  issn      = {1095-9203},
  number    = {5806},
  pages     = {1757-1761},
  url       = {http://dx.doi.org/10.1126/science.1133734},
  volume    = {314},
  journal   = {Science},
  month     = dec,
  publisher = {American Association for the Advancement of Science (AAAS)},
  year      = {2006},
}

@Article{Konig2007,
  author    = {K\"onig, Markus and Wiedmann, Steffen and Br\"une, Christoph and Roth, Andreas and Buhmann, Hartmut and Molenkamp, Laurens W. and Qi, Xiao-Liang and Zhang, Shou-Cheng},
  title = {{Quantum Spin Hall Insulator State in HgTe Quantum Wells}},
  doi       = {10.1126/science.1148047},
  issn      = {1095-9203},
  number    = {5851},
  pages     = {766-770},
  url       = {http://dx.doi.org/10.1126/science.1148047},
  volume    = {318},
  journal   = {Science},
  month     = nov,
  publisher = {American Association for the Advancement of Science (AAAS)},
  year      = {2007},
}

@Article{Hsieh2008,
  author    = {Hsieh, D. and Qian, D. and Wray, L. and Xia, Y. and Hor, Y. S. and Cava, R. J. and Hasan, M. Z.},
  title = {{A topological Dirac insulator in a quantum spin Hall phase}},
  doi       = {10.1038/nature06843},
  issn      = {1476-4687},
  number    = {7190},
  pages     = {970-974},
  url       = {http://dx.doi.org/10.1038/nature06843},
  volume    = {452},
  journal   = {Nature},
  month     = apr,
  publisher = {Springer Science and Business Media LLC},
  year      = {2008},
}

@Article{Hsieh2009,
  author    = {Hsieh, D. and Xia, Y. and Qian, D. and Wray, L. and Dil, J. H. and Meier, F. and Osterwalder, J. and Patthey, L. and Checkelsky, J. G. and Ong, N. P. and Fedorov, A. V. and Lin, H. and Bansil, A. and Grauer, D. and Hor, Y. S. and Cava, R. J. and Hasan, M. Z.},
  title = {{A tunable topological insulator in the spin helical Dirac transport regime}},
  doi       = {10.1038/nature08234},
  issn      = {1476-4687},
  number    = {7259},
  pages     = {1101-1105},
  url       = {http://dx.doi.org/10.1038/nature08234},
  volume    = {460},
  journal   = {Nature},
  month     = jul,
  publisher = {Springer Science and Business Media LLC},
  year      = {2009},
}

@article{Hasan2010,
  title = {{Colloquium: Topological insulators}},
  author = {Hasan, M. Z. and Kane, C. L.},
  journal = {Rev. Mod. Phys.},
  volume = {82},
  issue = {4},
  pages = {3045--3067},
  numpages = {0},
  year = {2010},
  month = {Nov},
  publisher = {American Physical Society},
  doi = {10.1103/RevModPhys.82.3045},
  url = {https://link.aps.org/doi/10.1103/RevModPhys.82.3045}
}

@Article{Fu2011TCI,
  title = {{Topological Crystalline Insulators}},
  author = {Fu, Liang},
  journal = {Phys. Rev. Lett.},
  volume = {106},
  issue = {10},
  pages = {106802},
  numpages = {4},
  year = {2011},
  month = {Mar},
  publisher = {American Physical Society},
  doi = {10.1103/PhysRevLett.106.106802},
  url = {https://link.aps.org/doi/10.1103/PhysRevLett.106.106802}
}

@article{Tanaka2012,
  title = {{Experimental realization of a topological crystalline insulator in SnTe}},
  volume = {8},
  ISSN = {1745-2481},
  url = {http://dx.doi.org/10.1038/nphys2442},
  DOI = {10.1038/nphys2442},
  number = {11},
  journal = {Nat. Phys.},
  publisher = {Springer Science and Business Media LLC},
  author = {Tanaka,  Y. and Ren,  Zhi and Sato,  T. and Nakayama,  K. and Souma,  S. and Takahashi,  T. and Segawa,  Kouji and Ando,  Yoichi},
  year = {2012},
  month = sep,
  pages = {800}
}

@article{Dziawa2012,
  title = {{Topological crystalline insulator states in Pb$_{1-x}$Sn$_x$Se}},
  volume = {11},
  ISSN = {1476-4660},
  url = {http://dx.doi.org/10.1038/nmat3449},
  DOI = {10.1038/nmat3449},
  number = {12},
  journal = {Nat. Mater.},
  publisher = {Springer Science and Business Media LLC},
  author = {Dziawa,  P. and Kowalski,  B. J. and Dybko,  K. and Buczko,  R. and Szczerbakow,  A. and Szot,  M. and Łusakowska,  E. and Balasubramanian,  T. and Wojek,  B. M. and Berntsen,  M. H. and Tjernberg,  O. and Story,  T.},
  year = {2012},
  month = sep,
  pages = {1023--1027}
}

@article{Wan2011WeylSemimetal,
  title = {{Topological semimetal and Fermi-arc surface states in the electronic structure of pyrochlore iridates}},
  author = {Wan, Xiangang and Turner, Ari M. and Vishwanath, Ashvin and Savrasov, Sergey Y.},
  journal = {Phys. Rev. B},
  volume = {83},
  issue = {20},
  pages = {205101},
  numpages = {9},
  year = {2011},
  month = {May},
  publisher = {American Physical Society},
  doi = {10.1103/PhysRevB.83.205101},
  url = {https://link.aps.org/doi/10.1103/PhysRevB.83.205101}
}

@article{Burkov2011WeylSemimetal,
  title = {{Weyl Semimetal in a Topological Insulator Multilayer}},
  author = {Burkov, A. A. and Balents, Leon},
  journal = {Phys. Rev. Lett.},
  volume = {107},
  issue = {12},
  pages = {127205},
  numpages = {4},
  year = {2011},
  month = {Sep},
  publisher = {American Physical Society},
  doi = {10.1103/PhysRevLett.107.127205},
  url = {https://link.aps.org/doi/10.1103/PhysRevLett.107.127205}
}

@article{Singh2012WeylSemimetal,
  title = {{Topological electronic structure and Weyl semimetal in the TlBiSe$_2$ class of semiconductors}},
  author = {Singh, Bahadur and Sharma, Ashutosh and Lin, H. and Hasan, M. Z. and Prasad, R. and Bansil, A.},
  journal = {Phys. Rev. B},
  volume = {86},
  issue = {11},
  pages = {115208},
  numpages = {7},
  year = {2012},
  month = {Sep},
  publisher = {American Physical Society},
  doi = {10.1103/PhysRevB.86.115208},
  url = {https://link.aps.org/doi/10.1103/PhysRevB.86.115208}
}

@article{Young2012DiracSemimetal,
  title = {{Dirac Semimetal in Three Dimensions}},
  author = {Young, S. M. and Zaheer, S. and Teo, J. C. Y. and Kane, C. L. and Mele, E. J. and Rappe, A. M.},
  journal = {Phys. Rev. Lett.},
  volume = {108},
  issue = {14},
  pages = {140405},
  numpages = {5},
  year = {2012},
  month = {Apr},
  publisher = {American Physical Society},
  doi = {10.1103/PhysRevLett.108.140405},
  url = {https://link.aps.org/doi/10.1103/PhysRevLett.108.140405}
}

@article{Armitage2018WeylDirac,
  title = {{Weyl and Dirac semimetals in three-dimensional solids}},
  author = {Armitage, N. P. and Mele, E. J. and Vishwanath, Ashvin},
  journal = {Rev. Mod. Phys.},
  volume = {90},
  issue = {1},
  pages = {015001},
  numpages = {57},
  year = {2018},
  month = {Jan},
  publisher = {American Physical Society},
  doi = {10.1103/RevModPhys.90.015001},
  url = {https://link.aps.org/doi/10.1103/RevModPhys.90.015001}
}

@Article{Hasan2017Review,
  author    = {Hasan, M. Zahid and Xu, Su-Yang and Belopolski, Ilya and Huang, Shin-Ming},
  title     = {{Discovery of Weyl Fermion Semimetals and Topological Fermi Arc States}},
  doi       = {10.1146/annurev-conmatphys-031016-025225},
  issn      = {1947-5462},
  number    = {1},
  pages     = {289},
  url       = {http://dx.doi.org/10.1146/annurev-conmatphys-031016-025225},
  volume    = {8},
  fjournal  = {Annual Review of Condensed Matter Physics},
  journal   = {Annu. Rev. Conden. Ma. P.},
  month     = mar,
  publisher = {Annual Reviews},
  year      = {2017},
}

@article{Slager2012,
  author = {Slager,  Robert-Jan and Mesaros,  Andrej and Juričić,  Vladimir and Zaanen,  Jan},
  title = {{The space group classification of topological band-insulators}},
  number = {2},
  volume = {9},
  pages = {98},
  ISSN = {1745-2481},
  url = {http://dx.doi.org/10.1038/nphys2513},
  DOI = {10.1038/nphys2513},
  journal = {Nat, Phys.},
  publisher = {Springer Science and Business Media LLC},
  year = {2012},
  month = Dec,
}

@article{Kruthoff2017,
  title = {{Topological Classification of Crystalline Insulators through Band Structure Combinatorics}},
  author = {Kruthoff, Jorrit and de Boer, Jan and van Wezel, Jasper and Kane, Charles L. and Slager, Robert-Jan},
  journal = {Phys. Rev. X},
  volume = {7},
  issue = {4},
  pages = {041069},
  numpages = {23},
  year = {2017},
  month = {Dec},
  publisher = {American Physical Society},
  doi = {10.1103/PhysRevX.7.041069},
  url = {https://link.aps.org/doi/10.1103/PhysRevX.7.041069}
}

@book{bir1974symmetry,
  title={{Symmetry and Strain-Induced Effects in Semiconductors}},
  author={Bir, G.L. and Pikus, G.E.},
  isbn={9780470073216},
  lccn={74014842},
  series={A Halsted Press book},
  url={https://books.google.com.br/books?id=38m2QgAACAAJ},
  year={1974},
  publisher={Wiley},
  address   = {New York},
}

\end{document}